\documentclass[lettersize,journal,twoside]{IEEEtran}
\usepackage{amsmath,amssymb,amsfonts}
\usepackage{algorithmic}
\usepackage{algorithm}
\usepackage{array}
\usepackage{textcomp}
\usepackage{stfloats}
\usepackage{url}
\usepackage{verbatim}
\usepackage{graphicx}
\usepackage{subcaption} 
\usepackage{cite}
\usepackage{xcolor}
\usepackage{color,soul}
\usepackage{makecell}
\usepackage{multirow}
\usepackage[colorlinks=true,
            linkcolor=blue,
            urlcolor=black,
            citecolor=blue,
            anchorcolor=blue]{hyperref}
\renewcommand{\arraystretch}{1.3} 

\ifodd 1
\newcommand{\CommentWong}[1]{\textcolor[rgb]{1,0,0}{[Wong comment: #1]}}
\newcommand{\CommentMush}[1]{\textcolor[rgb]{0,0,1}{[Mushfiqur comment: #1]}}

\else
\newcommand{\CommentWong}[1]{}
\newcommand{\CommentMush}[1]{}

\fi

\begin{document}

\title{Individualized Deepfake Detection Exploiting Traces Due to Double Neural-Network Operations}

\author{Mushfiqur~Rahman, 
Runze~Liu, 
Chau-Wai~Wong,~\IEEEmembership{Senior Member,~IEEE,}
and~Huaiyu~Dai,~\IEEEmembership{Fellow,~IEEE}%
\thanks{This work was supported in part by the US National Science Foundation (award number ECCS-2227499) \textit{(Corresponding author: Chau-Wai Wong}).}%
\thanks{Mushfiqur Rahman, Chau-Wai Wong, and Huaiyu Dai are with the Department of Electrical and Computer Engineering, NC State University, NC 27695 USA.}%
\thanks{Runze Liu is now an independent researcher. He conducted this research work he was with the Department of Electrical and Computer Engineering, NC State University, NC 27695 USA.}%
\thanks{The source code and dataset are available at 
\url{https://github.com/rmushfiqur2/deepfake_op_rel}.}
}

\maketitle

\begin{abstract}
In today's digital landscape, journalists urgently require tools to verify the authenticity of facial images and videos depicting specific public figures before incorporating them into news stories. Existing deepfake detectors are not optimized for this detection task when an image is associated with a specific and identifiable individual. This study focuses on the deepfake detection of facial images of individual public figures. We propose to condition the proposed detector on the identity of an identified individual, given the advantages revealed by our theory-driven simulations. While most detectors in the literature rely on perceptible or imperceptible artifacts present in deepfake facial images, we demonstrate that the detection performance can be improved by exploiting the idempotency property of neural networks. In our approach, the training process involves double neural-network operations where we pass an authentic image through a deepfake simulating network twice. Experimental results show that the proposed method improves the area under the curve (AUC) from 0.92 to 0.94 and reduces its standard deviation by 17\%. 
To address the need for evaluating detection performance for individual public figures, we curated and publicly released a dataset of $\sim$32k images featuring 45 public figures, as existing deepfake datasets do not meet this criterion.
\end{abstract}


\section{Introduction}
\label{sec:intro}
\begin{figure*}[!t]
    \centering
    \begin{subfigure}[b]{0.32\linewidth}
        \centering
        \includegraphics[width=\linewidth,trim={3.2cm 0.2cm 3.6cm 0.5cm},clip]{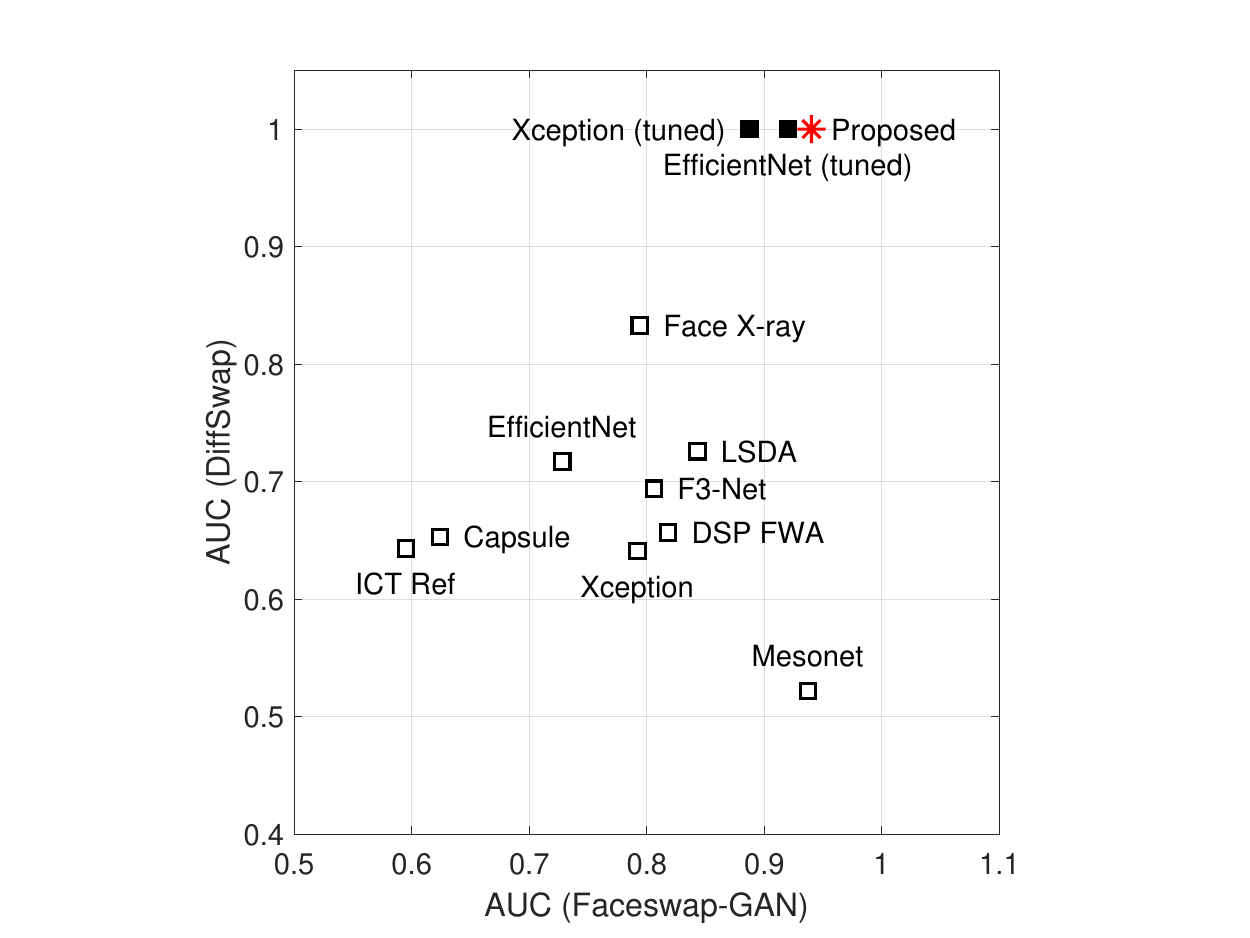}
        \caption{}
        \label{fig_intro2:image1}
    \end{subfigure}
    \begin{subfigure}[b]{0.32\linewidth}
        \centering
        \includegraphics[width=\linewidth,trim={6.0cm 2.0cm 8.8cm 1.8cm},clip]{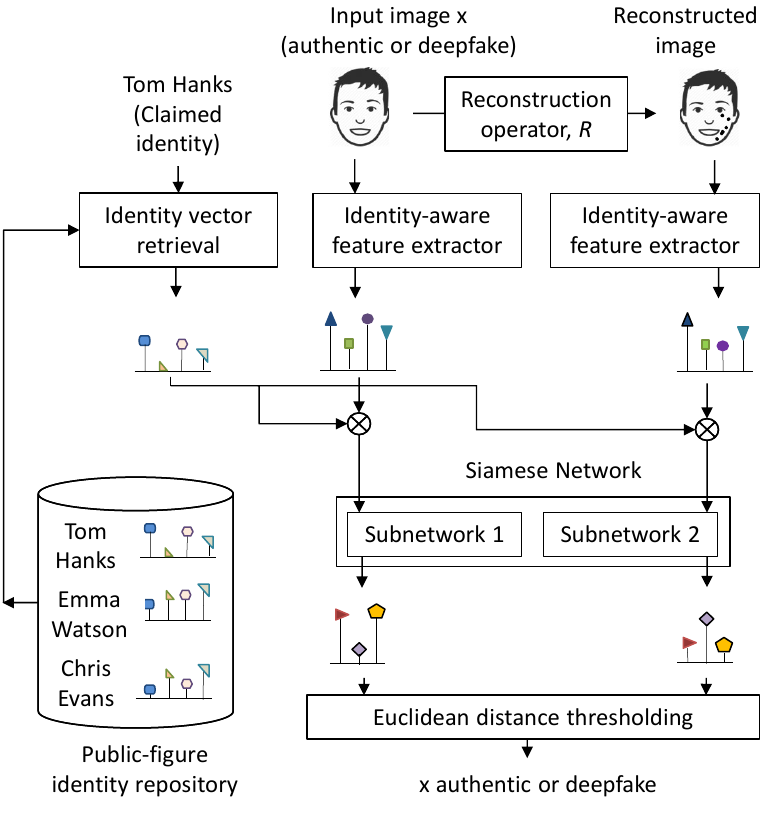}
        \caption{}
        \label{fig_intro:image1}
    \end{subfigure}
    \begin{subfigure}[b]{0.32\linewidth}
        \centering
        \includegraphics[width=\linewidth,trim={6.0cm 2.8cm 6.8cm 2.8cm},clip]{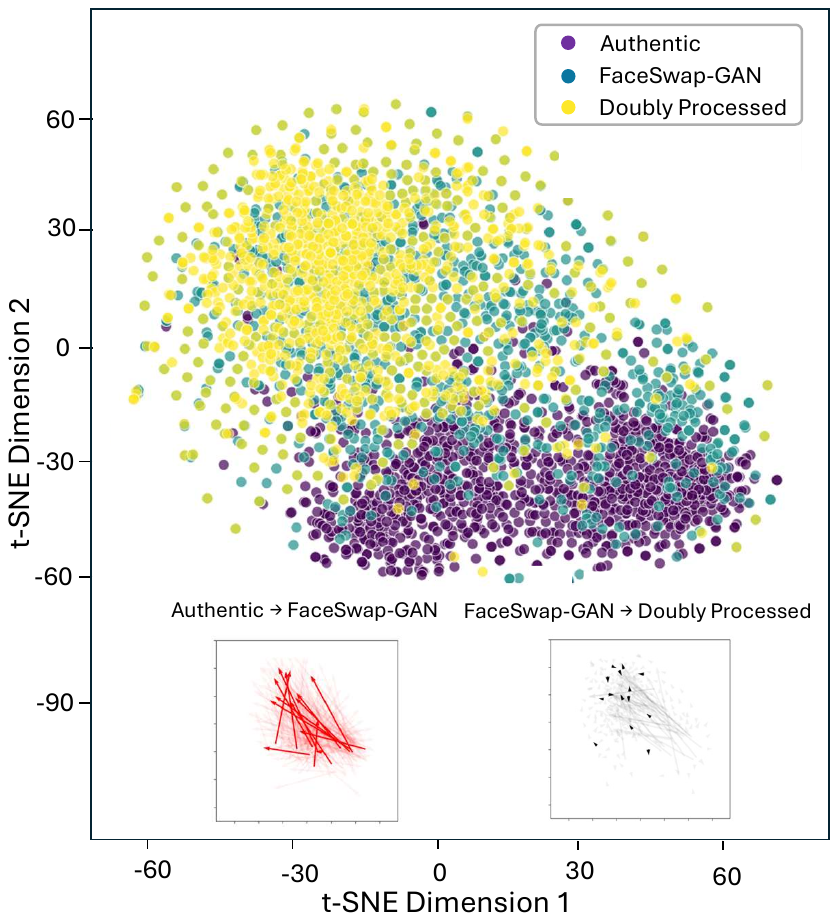} 
        \caption{}
        \label{fig_intro:image2}
    \end{subfigure}
    \caption{(a) Comparison of AUC performance for nine off-the-shelf deepfake detection methods (listed in TABLE~\ref{tab:summary3}), two fine-tuned methods, and the proposed method, evaluated on GAN-based Faceswap-GAN~\cite{fsg} deepfakes and diffusion-based DiffSwap~\cite{zhao2023diffswap} deepfakes. Square markers denote methods without finetuning (unfilled) and with finetuning (filled), while the star marker highlights the proposed method. Method names are labeled near their respective markers for better visualization. (b) The inference pipeline of the proposed individualized deepfake detector leveraging the near-idempotence property and identity conditioning. The identity conditioning is achieved by combining the identity-aware processing trace and the input identity vector. To leverage the idempotence property, the test image is passed through a reconstruction operator $R$. If the test image exhibits a marginal change in the observed amount of processing traces, the test image is considered ``deepfake"; if a significant change is observed, the image is considered ``authentic." (c) t-SNE visualization of authentic, deepfake, and doubly-processed images and their corresponding vector shifts in t-SNE feature space across the two transformations. Red arrows indicate the vector shifts for the first transformation, while black arrows represent shifts during the second transformation. The first transformation causes a significant change when a deepfake operator initially processes an authentic image. In contrast, the second transformation results in only minor shifts. For more details, please refer to Section~\ref{sec:veri_near_idem}.}
    \label{fig:diagram_proposed_testing}
    \vspace{-0.5cm}
\end{figure*}
A deepfake refers to a seemingly authentic image or video generated by a deep neural network. When it comes to human faces, a manipulation method may comprise reenactment, replacement, editing, and synthesis~\cite{mirsky2021creation}. While deepfakes can facilitate numerous appealing and advantageous applications, the act of replacing the face in a staged image or video with the face of a public figure can pose a serious threat to society. Given the continuous influx of deepfake videos on public platforms, journalists need to pay special attention to those that relate to significant public interest, such as those featuring celebrities or politicians~\cite{mirsky2021creation, durall2020watch}. The deepfake generation methods evolved with autoencoder-based approaches~\cite{GithubDeepfake}, GANs~\cite{fsg}, and diffusion models~\cite{rombach2022high}. The latest diffusion-based models, such as \cite{zhao2023diffswap, rombach2022high, saharia2022photorealistic}, can surpass GAN-based models in producing photorealistic images.
Nevertheless, even in the present day, autoencoder-based models remain threatening in terms of malicious use. This is due to the availability of several free, downloadable, and user-friendly applications built on autoencoder, such as FaceSwap~\cite{fs}, Faceswap-GAN~\cite{fsg}, DeepFaceLab~\cite{dflab}, and df~\cite{dfaker}. In this work, we focus on GAN-based Faceswap-GAN~\cite{fsg} and diffusion-based DiffSwap~\cite{zhao2023diffswap}.

Most deepfake detectors were built to detect the whole population of deepfake videos, i.e., deepfake videos of whatever identities are targeted. However, victims of deepfakes are most often public figures and their deepfake videos are more detrimental due to their widespread public exposure.
In this work, we propose a deepfake image detection system customized for individual subjects. Our theory-driven simulations suggest that identity conditioning on deepfake detection tends to exhibit advantages in more challenging detection tasks. As our experimental results will show, the existing tools for deepfake face detection that encompass the whole population may work suboptimally for a specific public figure. The proposed detector for specific individuals is especially useful for journalism. For example, before reporting news based on an image of a public figure of unknown authenticity, a journalist can apply the proposed detection tool to determine its authenticity.

Our approach to deepfake detection draws inspiration from a series of studies leveraging the near-idempotence property of an operation. This method has been particularly effective in various image forensics tasks, including double JPEG compression detection, unknown video codec identification, and source camera identification ~\cite{huang2010detecting,yang2014effective,bestagini2012video,deng2011source,milani2014demosaicing}. In these studies, researchers leverage the near-idempotence of a respective operation, such as certain type of JPEG compression, video compression, or color demosaicing algorithm. The strict idempotence property asserts that an \textit{idempotent} operation, $g(\cdot)$, results in no change to $g(x)$ when it is applied iteratively, i.e., $g(g(x)) = g(x)$. Using slightly different terminology, if $g(g(x))$ approximately equals $g(x)$, the operation is \textit{nearly idempotent}. In many detection problems of multimedia forensics, the nearly idempotent nature of a forgery method allows an analyst to apply the forgery operation multiple times and observe the changes to determine whether the input was forged for the first time, i.e., input forged for more than once will exhibit minimal changes.

In this work, we demonstrate that near-idempotence is also applicable to the neural network-based Faceswap-GAN~\cite{fsg} and DiffSwap~\cite{zhao2023diffswap}, as demonstrated in Fig.~\ref{fig:diagram_proposed_testing}(c). To explore this, we emulate a potential deepfake operation that an attacker might employ, utilizing publicly available data of a public figure and employing a neural network architecture to replicate the functionality of a deepfake generation tool. Referring to it as the reconstruction operator \( R \), Fig.~\ref{fig:diagram_proposed_testing}(b) illustrates the inference pipeline of the proposed detector. We feed a test image into the emulated deepfake generator. The expected change in the image due to this operation is dependent on whether the image has undergone a similar operation before. If the image is a deepfake, the near-idempotence property ensures that the change will be minimal. From the standpoint of the deepfake feature extractor, a deepfake image will exhibit processing traces both before and after the operation, leading to subtle observed changes. Conversely, an authentic image without the deepfake operation lacks any processing traces of the neural network, resulting in a significant observable change. The contributions of this paper are threefold.
\begin{itemize}
    \item We propose to use the near-idempotence property of neural networks for deepfake face detection, introducing a distinct direction of improvement compared to the state of the art. The idempotence-driven approach can potentially complement existing methods.
    \item We demonstrate that identity conditioning can significantly improve the deepfake detection performance over
    the state-of-the-art end-to-end CNN classifiers.
    \item Our detector can focus on specific individuals. Individualized detectors are better suited for journalism. 
\end{itemize}

The remainder of this paper is organized as follows. Section~\ref{sec:related_work} discusses the existing literature on deepfake generation, detection, and approaches related to the proposed method. Section~\ref{sec:threat_model} introduces the threat model, while Section~\ref{sec:method} presents the proposed deepfake detection method based on near-idempotence and identity conditioning. Section~\ref{sec:experimental_results} showcases the experimental results, followed by Section~\ref{sec:discussion}, highlighting the key findings of this work. Finally, Section~\ref{sec:conclusion} concludes the paper.
\section{Related Work}
\label{sec:related_work}
\subsection{Generation of Deepfake Faces}
Early methods of face-swapping, such as Bitouk et al.~\cite{bitouk2008face}, were limited to using two images of two particular persons with similar poses. The images were first aligned with the help of landmark detection, then cropped and postprocessed, including color correction. Subsequent researchers~\cite{cheng20093d} improved those with a 3-D facial model from the source video. The next advancement emerged after the proposal of a deep-learning-based face-swapping architecture~\cite{GithubDeepfake} built upon one shared encoder and two individual decoders. Faceswap-GAN~\cite{fsg} is the GAN improvement over faceswap~\cite{GithubDeepfake}, where the performance of shared encoder and individual decoders further improve as a result of the GAN's internal interplay mechanism between the generator and discriminator. However, the architectures proposed in \cite{GithubDeepfake, fsg} can only swap faces between the two identities involved in training. Researchers have proposed identity-agnostic architectures that decouple identity extraction from attribute extraction \cite{bao2018towards, nirkin2019fsgan, natsume2019fsnet, natsume2018rsgan, li2019faceshifter}. Recent works~\cite{song2023robustness, ivanovska2024vulnerability} demonstrate that denoising diffusion models~(DDMs) can significantly reduce the performance of deepfake detectors trained on images not generated by DDMs. For example, DiffSwap~\cite{zhao2023diffswap} considers face swapping as a conditional image inpainting task, where the denoising network is conditioned on the identity features of the source image and the facial landmarks of the target image. DiffFace~\cite{kim2022diffface} employs a diffusion model with a facial guidance mechanism incorporating three distinct control components for identity, semantic features, and gaze for maintaining consistent pose and facial attributes.

\subsection{Protection Against Deepfakes}
Researchers have explored a variety of techniques for deepfake detection. Some exploit the artifacts of synthetic videos, such as the absence of eye blinking~\cite{li2018ictu}, inconsistency in head pose~\cite{yang2019exposing}, disparities in color components~\cite{li2018detection}, and inconsistency between inner face and outer face~\cite{dong2022protecting}. Some other researchers opt for a complete data-driven approach by using either an end-to-end convolutional neural network (CNN) structure~\cite{wang2020cnn} or a combined CNN with a recurrent neural network (RNN)~\cite{guera2018deepfake}. Researchers have also exploited processing traces left by the neural networks for deepfake detection. The researchers exploited the features such as spatial domain local convolutional features~\cite{guarnera2020deepfake, afchar2018mesonet, nguyen2019use, chollet2017xception, li2020face, bonettini2020video, yan2024transcending} and spectral distortion or upsampling artifacts in the frequency domain~\cite{durall2020watch, frank2020leveraging, li2018exposing, qian2020thinking}.

Instead of detecting deepfake videos for the whole population, the characteristics of a specific person have also been exploited. 
Agarwal et al.~\cite{agarwal2019protecting} targeted deepfake videos of a specific individual by capturing speaking patterns.
Cozzolino et al.~\cite{cozzo} proposed to learn the temporal features of how a specific person moves and talks.
Dong et al.~\cite{dong2022protecting} calculated $\ell^2$ distance between the computed identity vector from the inner face and the expected identity vector drawn from a reference set of identity vectors. In this work, we extract the deepfake traces conditioned on the identity.

\subsection{Idempotency as a Multimedia Forensics Tool}
In multimedia forensics, one way to detect counterfeiting is to exploit the near-idempotence property, i.e., the minor changes caused by the repetitive application of adversarial operations. 
It shares the same spirit of the law of diminishing returns, a widely used concept in economics~\cite{brue1993retrospectives,spillman1923application}.
The detection of double JPEG compression, source camera identification, and video codec identification are three exemplary applications of the near-idempotence property.
The ratio of stable image blocks has been used by researchers to detect the number of prior JPEG compressions~\cite{lai2013block,carnein2015forensics}.
Huang et al.~\cite{huang2010detecting} found that the number of dissimilar JPEG coefficients between two subsequent JPEG compression decreases monotonically. 
Bestagini et al.~\cite{bestagini2012video} detected unknown video encoding by recompressing a video with each of the candidates.
For source camera identification, the researchers have leveraged the near-idempotence property of an auto-white balancing method~\cite{deng2011source} and that of color demosaicing strategy~\cite{milani2014demosaicing}.
In economics, the law of diminishing returns states that additional inputs to a fixed amount of identical inputs increase productivity at a decreasing rate~\cite{brue1993retrospectives}. If the additional inputs are considered repetitive operations, then the law of diminishing returns may be regarded as near-idempotence.
In this study, we show that the near-idempotence property of neural networks assists in deepfake image detection.

\subsection{Unsupervised Pretraining}
Unsupervised pretaining has been proposed for feature extraction for many tasks of computer vision. Chen et al.~\cite{chen2020big} found that larger networks, for example, larger ResNet, pretrained in an unsupervised manner followed by supervised training with only $10\%$ of labeled data can outperform fully supervised networks for general computer vision tasks. Newell and Deng \cite{newell2020useful} showed that pretrained networks are more advantageous in low data regimes compared to ubiquitous data. Their results suggest that pretrained networks should be tested on diverse downstream tasks. Bulat et al.~\cite{bulat2022pre} proposed task-agnostic self-supervised pretraining on in-the-wild facial data for representation learning. Zheng et al.~\cite{zheng2022general} proposed weakly supervised facial representation learning using vast facial images available on the web with linguistic descriptions. In this work, we fine-tune the facial features from Bulat et al.~\cite{bulat2022pre} to learn the deepfake traces.
\section{Threat Model}
\label{sec:threat_model}
In this work, we consider an attacker who is smart enough to find and use open-source face-swapping software such as~\cite{fsg, fs, GithubDeepfake, zhao2023diffswap} on the facial images from the publicly available videos of a public figure. More specifically, we consider Faceswap-GAN~\cite{fsg} and DiffSwap~\cite{zhao2023diffswap} as potential methods that the attacker can use. The attacker is free to use any public or private videos of a second person to depict a story and convince the public of the involvement of a targeted public figure. For example, the attacker can record prearranged videos at a professional studio and later replace the actor's face with that of a public figure. The attacker can harvest videos of the public figure from multiple sources, including social media, news channels, movies, and YouTube. Different sources of videos offer varied image quality, compression levels, and processing histories. For example, public interview videos of a public figure available on YouTube are expected to be less edited than video clips from movies. In our proposed detection method, we assume that we, as forensic analysts, have access to the various sources of public figure videos, but we do not know exactly from what source the attacker took videos for deepfake generation. For example, the attacker can use videos from social media, where we will only use public interview recordings of that public figure to train the neural network-based detector.

\section{Proposed Detector via Near-Idempotence and Identity Conditioning}
\label{sec:method}
\begin{figure*}[!t]%
    \centering
		\vspace{-0mm}
    \includegraphics[width=\linewidth,trim={0.1cm 2.75cm 1.7cm 4.2cm},clip]{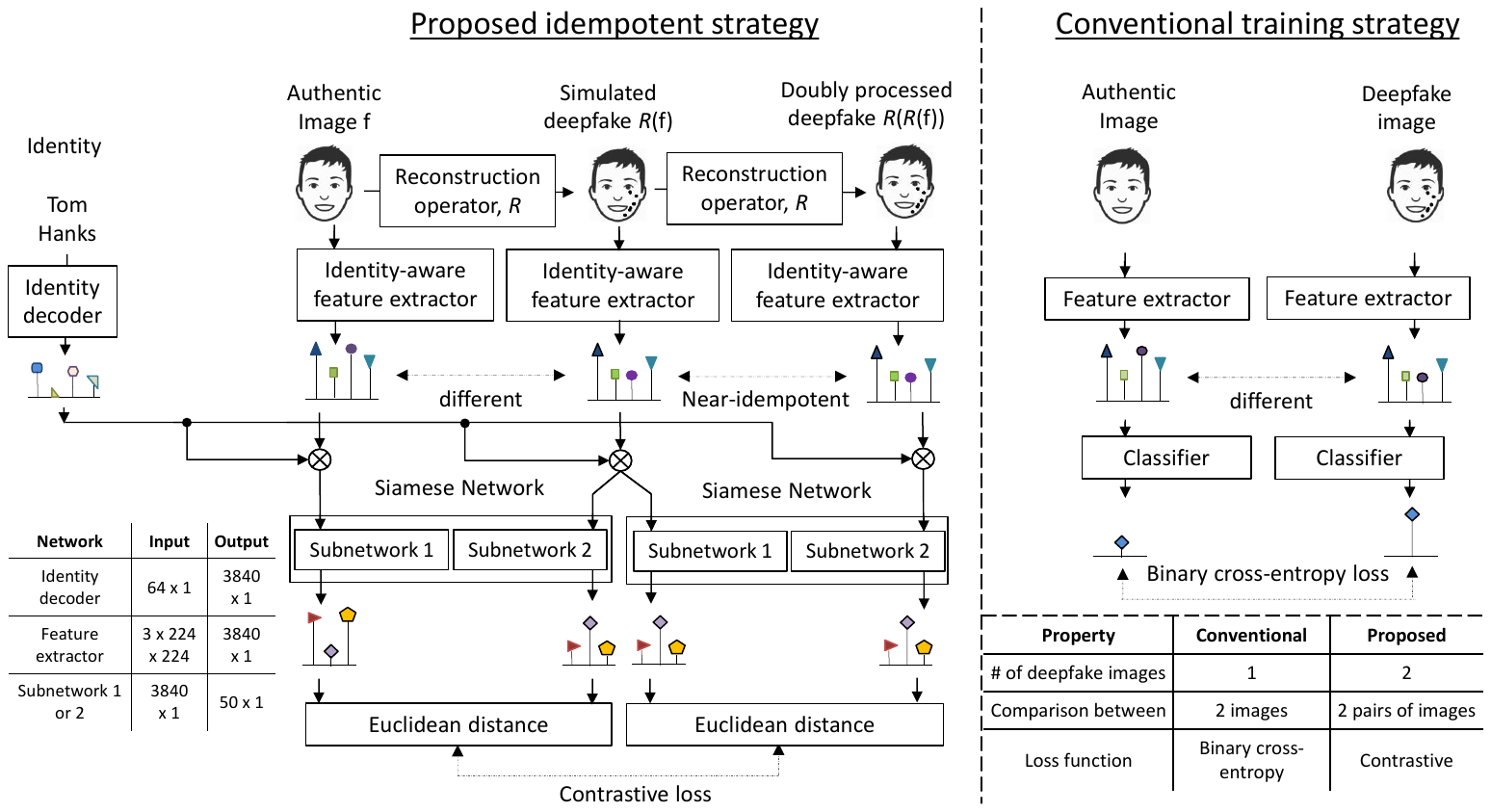}
    \hspace{4mm}
    \vspace{-1mm}
    \caption{The training pipeline of the proposed deepfake detector leveraging the near-idempotence property of the deepfake generator. A side-by-side comparison with conventional deepfake detectors is also shown. In the proposed method, an authentic image is passed through a deepfake simulating network or reconstruction operator twice. Due to the near-idempotence property, the features for the first and the second outputs will be nearly identical. The features are obtained from an identity-aware feature extractor that is trained separately. We freeze the feature extractor network and train a Siamese network and an identity decoder to increase the Euclidean distance between the first pair (consisting of the authentic image and the first output image) and to decrease the Euclidean distance between the second pair (consisting of the first and the second output images).}
	\vspace{-2mm}
    \label{fig:diagram_architecture_introduction}
\end{figure*}

In the challenge of identifying deepfake faces for public figures, we confront an image of unknown authenticity, claimed to be a specific public figure. Our approach to addressing this problem makes use of the extensive collection of authentic images or videos of the said public figure from YouTube.
The training process of our proposed deepfake detector is depicted in Fig.~\ref{fig:diagram_architecture_introduction}, and the inference pipeline is shown in Fig.~\ref{fig:diagram_proposed_testing}(b).
Our proposed detector has four distinct components. First, the reconstruction operator is a neural network operation that stimulates the deepfake generation operation for a public figure. We found this operation nearly idempotent. Second, the feature extractor is finetuned with a teacher network and is able to capture the identity information while extracting the features. Third, the identity decoder takes as input the explicit identity, i.e., the index of the public figure, and learns as a constant identity vector that arguments the feature space. It contains the necessary person-specific information of that public figure and, when combined with the identity-aware feature, can effectively compute the deepfake features conditioned on identity. Fourth, the Siamese network serves as the ultimate binary classification block in the proposed architecture. It learns to extract the features linked to the idempotency of the deepfake operation. It produces a larger distance before and after reconstruction for a test authentic image and a smaller distance for a test deepfake image.
\subsection{Reconstruction Operator and Idempotence-Driven Detection}
\label{sec:recon_operator}
We employ a dedicated reconstruction operator $R$ for each public figure as shown in Fig.~\ref{fig:diagram_proposed_testing}(b) and Fig.~\ref{fig:diagram_architecture_introduction}.
When the original image is authentic, the first operation generates a deepfake image, and the second operation produces a doubly processed deepfake. We verified experimentally that the reconstruction operator $R$ serves as a reliable approximation of a specific type of deepfake generation tool, such as FaceSwap-GAN~\cite{fsg}, and that the deepfake generation process is nearly idempotent. In this context, the distance between a deepfake image and its corresponding doubly processed deepfake tends to be close to zero. This characteristic is leveraged in the training and inference system.

\begin{figure}[!t]%
    \centering
		\vspace{-0mm}
	\subfloat[]{\includegraphics[width=0.48\linewidth]{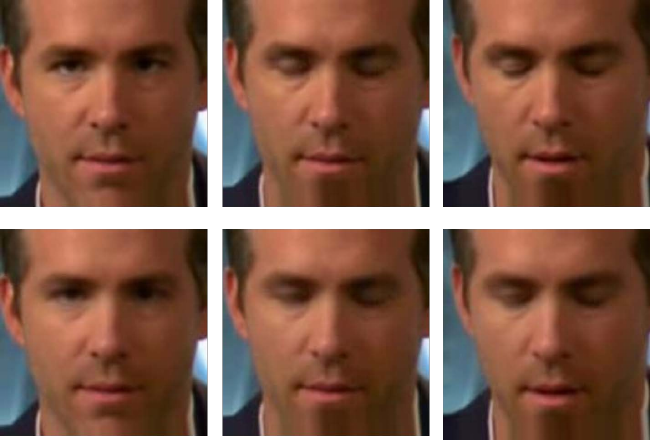}} 
	\hspace{2mm}
    \subfloat[]{\includegraphics[width=0.48\linewidth]{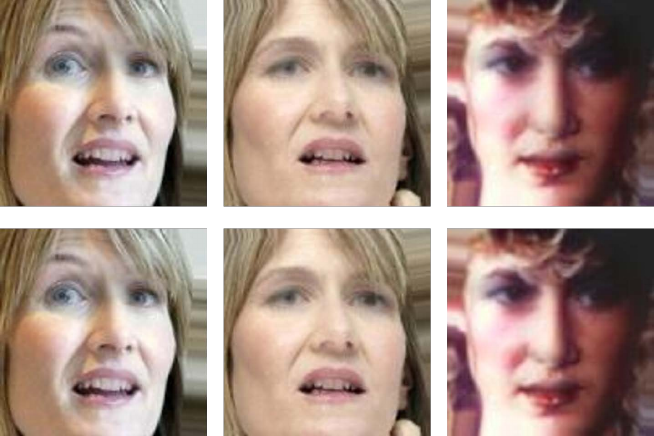}} 
    \vspace{-2mm}
    \caption{(a) Facial regions from raw images (first row) and reconstructed images (second row). The reconstructed images are singly processed. (b) Facial regions from deepfake images (first row) and reconstructed images (second row). The reconstructed images are doubly processed. The reconstruction models trained with images from the same person result in good visual quality for both raw and deepfake images.}
	\vspace{-2mm}
    \label{fig:reconstructed_frames}
\end{figure}

The next consideration is how to obtain the identity-specific reconstruction operator. For each public figure within our scope, we accumulate numerous images of that public figure and train a neural network based on an autoencoder utilizing the encoder and decoder architecture from FaceSwap-GAN~\cite{fsg}. This network learns the facial characteristics of the public figure, and when given a facial image of that public figure, it can reproduce approximately the same image as the output. Since the objective of this network is to replicate the input facial image of an identity, we refer to the resulting operator as the reconstruction operator or emulated deepfake generator. Some examples of reconstructed images are shown in Fig.~\ref{fig:reconstructed_frames}.

The reconstruction operator $R$ exhibits near-zero changes to a deepfake image due to the near-idempotence. Consequently, the feature level Euclidean distance between the two is expected to be small. On the other hand, an authentic image and its corresponding processed image will be substantially different as the operation leaves discernible traces in the processed image. Considering the capability of our deepfake feature extractor (see Section~\ref{sec:id-aware-feature-extractor}) to detect these traces, the features will exhibit significant dissimilarity, resulting in a higher distance compared to the deepfake scenario.

Based on the above considerations, the initial problem of detecting whether an image is authentic or deepfake is now reframed as evaluating the change of the image in the feature space through the reconstruction operation. When this change, quantified as the Euclidean distance, approaches zero, the image is classified as a deepfake; otherwise, it is considered authentic. Denoting the input image by $\text{f}$, the reframed problem is to evaluate whether $\text{f}$ and $R(\text{f})$ are the same or not, where $R$ is our reconstruction operator. Treating $\text{f}$ and $R(\text{f})$ as two inputs, we note that the Siamese network~\cite{bromley1993signature} is a powerful approach for discerning similarity or dissimilarity between two inputs. Our use of the Siamese network will be discussed in Section~\ref{sec:contrastive_learning}.

\subsection{Identity-Aware Feature Extractor}
\label{sec:id-aware-feature-extractor}

\subsubsection{Motivation}
Conventional deepfake feature extraction network $B(\cdot)$ extracts the deepfake features $B({\rm f})$ for a test image ${\rm f}$ ignoring the person identity ${\rm I}$~\cite{bonettini2020video,dang2020detection,zhao2021multi} or considers the identity features irrelevant to forgery detection~\cite{guo2023controllable, yan2023ucf}. Our work found that the identity-aware feature, $B^{'}({\rm f})$, which extracts identity information in addition to the deepfake features, is more effective for deepfake detection. This may be explained by the fact that a distinct extracted feature may not be equally distinguishable for every identity for the classification. If a feature extractor does not allow the passing of the identity information, the later network can not learn the statistics of the features individually for each identity. This will be limited to learning the average pattern.
Such average distributions of the features will lead to the error probability of the Bayesian classifier as follows:
\begin{equation}
\begin{aligned}
P_\text{e}^{\text{com}} = \mathbb{P}(H_0) \, \mathbb{P}\left(C\!=\!1 \mid H_0\right) + \mathbb{P}(H_1) \, \mathbb{P}\left(C\!=\!0 \mid H_1\right).
\end{aligned}
\end{equation}
where $\mathbb{P}(\cdot)$ is the probability measure, $H_0$ and $H_1$ are two hypotheses, $C$ is the predicted class. 
On the other hand, if the feature extractor allows passing the identity, the later network can distinguish the features for each identity separately.
Knowing the distributions of the features for each identity separately will lead to the error probability:
\begin{equation}
\begin{aligned}
P_\text{e}^{\text{ind}} \!=\! \frac{1}{N}\sum_{\text{I}\in\mathbb{I}} &\,\mathbb{P}\left(H_0 \mid \text{I}\right) \, \mathbb{P}\left(C\!=\!1 \mid H_0,\text{I}\right)\\
+&\,\mathbb{P}\left(H_1 \mid \text{I}\right) \, \mathbb{P}\left(C\!=\!0 \mid H_1,\text{I}\right),
\end{aligned}
\end{equation}
where $\mathbb{I}$ is the set of all identities.
In Appendix~\ref{sec:advantage_identity_aware} of the supplementary document, we showed that the latter identity-conditioning approach is more powerful in reducing classification error. We conducted a performance comparison between two methods through theory-driven simulations, demonstrating that $P_\text{e}^{\text{ind}}$ tends to be lower (better) than $P_\text{e}^{\text{com}}$. Furthermore, we observed that the gain of $P_\text{e}^{\text{ind}}$ over $P_\text{e}^{\text{com}}$ is more significant when the deepfake traces for individuals are more unique, and the detection problem is intrinsically more difficult.

\subsubsection{Training} 
To make the feature extraction network identity-aware, we use a neural network such that the earlier layers extract identity-aware features along with other features, and the later layers extract deepfake traces. We use a learned facial representation, trained by Bulat et al.~\cite{bulat2022pre} as the starting point of training $B^{'}({\cdot})$.
Their trained network has an architecture of ResNet. For extracting deepfake features, we tune the portion of the network after the ``conv4" block. 

We reused the model and initial weights from Bulat et al.~\cite{bulat2022pre} for the following three reasons. First, having an existing network that lets personal identity pass through makes our task easier to additionally learn the deepfake traces. In comparison, training a network simultaneously for personal identity and deepfake detection would require joint training of two downstream tasks, which is harder. Second, a deeper network trained with unlabelled data is less biased to any specific portion of the dataset~\cite{chen2020big}. Bulat et al.~\cite{bulat2022pre} pretrained the ResNet architecture with ${\sim}10$ million facial images. Consequently, the initial layers of the network are anticipated to learn a robust representation of features, including the identity. The network is also tested over multiple downstream tasks, and therefore, it is a good candidate for extracting facial features~\cite{newell2020useful}.
Third, according to Newell and Deng~\cite{newell2020useful}, there is an advantage in unsupervised pretraining with unlabeled data when the labeled finetuning dataset is small, which aligns with our labeled training dataset.

\begin{figure}[!t]%
    \centering
		\vspace{-0mm}
    \includegraphics[width=\linewidth,trim={2.3cm 0.15cm 11.2cm 2.1cm},clip]{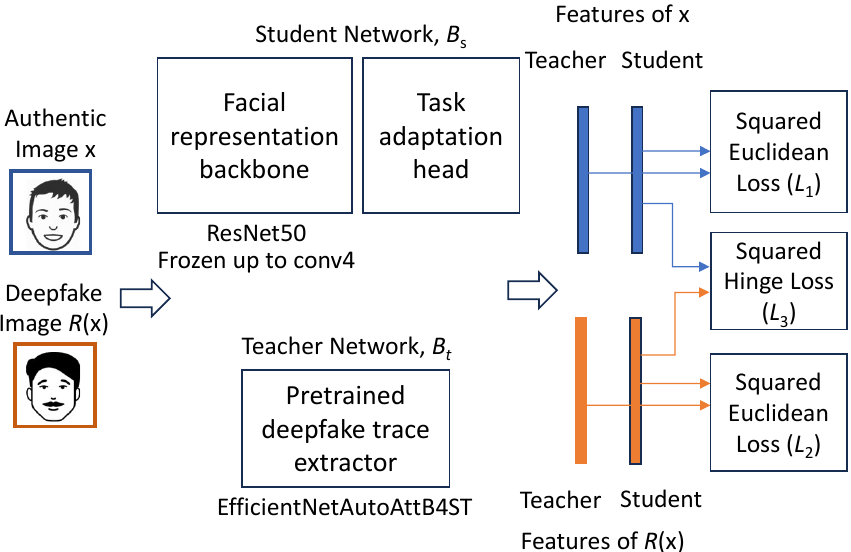}
    \hspace{4mm}
    \vspace{-1mm}
    \caption{Backbone network training for identity-aware deepfake feature extraction. An authentic and deepfake image pair is passed through the teacher and student networks. The teacher network passes down the deepfake trace knowledge to the student network through loss functions $L_1$ and $L_2$. The loss function $L_3$ increases the feature distance between the authentic and the deepfake image.}
	\vspace{-2mm}
    \label{fig:backbone-tuning}
\end{figure}

The training for the backbone network $B^{'}(\cdot)$ is depicted in Fig.~\ref{fig:backbone-tuning}. The input is an image pair consisting of an authentic image and its corresponding deepfake, generated using a deepfake generation tool. The input is passed through a student network $B_{\rm s}$ and a teacher network $B_{\rm t}$ in parallel. The student network is composed of the pretrained facial representation learning backbone~\cite{bulat2022pre} and a concatenated task adaptation head for learning the deepfake traces. The layers after the ``conv4" block of the pretrained backbone and the task adaptation head are the tunable portions of the student network. We then utilize the EfficientNetAutoAttB4ST~\cite{bonettini2020video} as the teacher network to distill the knowledge for learning the deepfake traces. To adapt the deepfake traces based on personal identity, we add a loss function $L_3$ that contrasts the learned traces of a deepfake and its corresponding authentic image in addition to the knowledge distillation losses $L_1$ and $L_2$.
Given the authentic facial image of identity $\rm I$, $\rm f_{auth}$, and its corresponding deepfake image $\rm f_{df}$,
the loss terms are defined as follows:
\begin{subequations}
\begin{align}
L_1 &= D^2\Big(B_{\rm t}({\rm f_{auth}}),B_{\rm s}(\rm f_{auth})\Big), \\
L_2 &= D^2\Big(B_{\rm t}({\rm f_{df}}),B_{\rm s}(\rm f_{df})\Big), \\
L_3 &= \Big[\max \Big(0, m_\text{h} \! - \! D\big(B_{\rm s}({\rm f_{auth}}),B_{\rm s}({\rm f_{df}})\big)\Big) \Big]^{2},
\end{align}
\end{subequations}
where $D(\cdot)$ is the Euclidean distance and $m_\text{h}$ is the margin of the hinge loss.
The three loss terms are combined as $\alpha(L_1 + L_2) + \beta L_3$, with hyperparameters $\alpha$ and $\beta$.
$L_1$ and $L_2$ contribute to the knowledge distillation for learning the deepfake traces, and $L_3$ contributes to learning the deepfake traces according to identity.
\subsection{Identity Decoder for Feature Conditioning}
Our identity decoder is a single-layer fully-connected neural network that maps the one-hot-encoded index of a public figure to the feature space generated by our feature extractor. We combine the output of the identity decoder with the output of the identity-aware feature extractor that contains the joint information of the deepfake feature and identity. The extra marginal information provided by the identity decoder can have the effect of conditioning the identity-aware feature in a similar spirit to the Bayes rule.

\subsection{Contrastive Learning}
\label{sec:contrastive_learning}
The Siamese network contains two identical subnetworks that process the two inputs parallelly. The subnetworks learn a manifold for each of the inputs, adopting contrastive loss that allows powerful discrimination between the two inputs. In our work, we designed each of the subnetworks as a single-layer neural network that takes as input the features of the corresponding image and outputs a vector of length $50$. We experimentally verified that this length is enough to discriminate between the two cases. Let us call the two subnetworks of the Siamese network $S_{n_1}$ and $S_{n_2}$, where the first one processes the features of $\text{f}$ and the second one processes the features of $R(\text{f})$.
We used contrastive loss \cite{hadsell2006dimensionality} to train the Siamese network as follows:
\begin{equation}
\begin{split}
L\big(\text{f}, \! R(\text{f}), \! Y\big)\!=\!(1 \!- \! Y)D^2_{S_n}{+}Y \! \left[\max \left(0, m \! - \! D_{S_n}\right) \right]^{2},
\end{split}
\end{equation}
\noindent where $D_{S_n}$ is the Euclidean distance between the processed manifolds, i.e., $D_{S_n} = \left\|S_{n_1}\left(X_{1}\right)-S_{n_2}\left(X_{2}\right)\right\|_{2}$, $X_1$ is the identity-conditioned features of $\text{f}$, $X_2$ is the identity-conditioned features of $R(\text{f})$, $m>0$ is a margin, and $Y \in \{0, 1\}$ is the known binary label of $\text{f}$, i.e., is $1$ if $\text{f}$ authentic, and $0$ otherwise.
We learned the weights of the identity decoder and the two subnetworks of the Siamese network using this loss function. Additionally, in contrast to the standard Siamese network, we decoupled the weights of the two subnetworks, $S_{n_1}$ and $S_{n_2}$, similar to CLIP~\cite{radford2021learning}, resulting in performance enhancement.
\section{Experimental Results}
\label{sec:experimental_results}
One key difference between our proposed method and the existing literature is the use of the near-idempotence property. This section validates the deepfake operation's near-idempotence property and experimentally demonstrates the performance gains leveraging this property.

\subsection{Dataset Curation}
The deepfake literature encourages cross-dataset evaluations, as they reveal a significant performance drop compared to in-dataset evaluations~\cite{chen2022self}. 
To conduct the cross-dataset evaluation with identity conditioning, we will need two separate datasets containing facial images of the same set of identities. However, the identity information is not included in the existing public deepfake detection datasets. For example, DFDC~\cite{dolhansky2020deepfake}, DFD~\cite{dfd}, and Deeper Forensics~\cite{jiang2020deeperforensics} do not explicitly mention identity information associated with the videos. This makes it difficult to find the same persons from another dataset, which would be necessary to perform the cross-dataset evaluation of individualized deepfake detection. To address this challenge, we curated a dataset with identities and a predefined train--test split, where the training and testing subsets are drawn from two different sources. Using our curated dataset, we report only the cross-dataset evaluation results. 

Our curated dataset contains 32,000 facial images of $45$ public figures sourced from Celeb-DF~\cite{Celeb_DF_cvpr20} for the training subset and from the cross-age facial image dataset~\cite{chen14cross} of the same public figures for the test dataset. We have publicly released the dataset, which contains authentic and deepfake images of public figures, their names, and a predefined train--test split.
For the training subset, we use real videos from the Celeb-DF dataset~\cite{Celeb_DF_cvpr20}, which is a popular deepfake detection dataset of 59 public figures. We sample frames from the videos at $5$ frames per second~(fps), and detect faces from the videos using the MTCNN~\cite{zhang2016joint} face detection network. For each individual~$i$, we have facial images $f^{\text{cdf}}_{i,j,k}$ from the $j$th authentic Celeb-DF video, where $j \in \{1,...,10\}$, $k \in \{1,...,N_f\}$, and $N_f$ is the number of the frames extracted from the video.

Examining multiple candidate datasets, we narrowed it down to the CACD~\cite{chen14cross} dataset for cross-dataset evaluation. CACD~\cite{chen14cross} contains cross-age facial images of 2,000 public figures with an overlap of 45 public figures with Celeb-DF. From CACD, we have authentic images $f^{\text{cacd}}_{i,j,k}$ for the $i$th identity and $j$th available age group of that identity, where $j \in \{1,...,5\}$, $k \in \{1,...,N_i\}$, and $N_i$ is the number of the images available for that age group. To generate deepfake faces for the testing subset, for each individual~$i$, we choose another identity $m$ from the database of 2,000 persons and then generate deepfake images using Faceswap-GAN~\cite{fsg} and DiffSwap~\cite{zhao2023diffswap} with the facial images $f^{\text{cacd}}_{i,j,k}$ and $f^{\text{cacd}}_{m,j,k}$.

\subsection{Experimental Setup}
Our proposed method had two stages of training. In the first stage, we trained the identity-aware feature extractor. For this training, we resized the facial images to 224-by-224 and used random cropping and random horizontal flipping for image augmentation. As shown in Fig.~\ref{fig:backbone-tuning}, we used a pair of images for the backbone training. We enforced identical cropping within the same pair, which consisted of an authentic image and its deepfake. We used the minibatch SGD optimizer from PyTorch with a learning rate of $10^{-3}$. The hinge loss margin $m_\text{h}$ was set to 50. During the first half of the iterations, a specific pair of values for \((\alpha, \beta)\) was used, which was then switched to a different pair for the second half. In each interval, the values were varied within the \([0,1] \times [0,1]\) region with a grid resolution of \(0.25 \times 0.25\).
The best results were obtained by setting \((\alpha, \beta) = (1, 0)\) during the initial 1,500 epochs of training, followed by a modification to \((\alpha, \beta) = (0, 1)\) for the subsequent 1,500 epochs.
In the second stage, we trained the Siamese network and the identity decoder. For this training, we used Adam optimizer, and the contrastive loss margin $m$ was $2$, and the learning rate was determined by the grid search within the range of $[10^{-6}, 10^{-5}]$ with a step size of $10^{-6}$.

For face reconstructor training, we separated the facial images from the last five videos $f^{\text{cdf}}_{i,j,k}, j \in \{6,\dots,10\}$ of the Celeb-DF dataset.
For the final classification network training, we randomly selected facial images from one video $f^{\text{cdf}}_{i,j,k}, j \in\{1,\dots,5\}$ as the validation set and facial images from other four videos as the training set. We repeated this process four times to ensure the results would be statistically stable. As for the test set, we used all of the real and face-swapped images that we generated from CACD.
In each training session, the neural network with the smallest validation loss was chosen as the final network for the test set.

\subsection{Baseline Algorithms Selection}
Our proposed double neural network-based detector is intended to boost a baseline algorithm. When selecting baseline algorithms, we ensured that they provided reliable numerical performance and also functioned as effective feature extractors for integration with our proposed detector. We initially picked nine state-of-the-art deepfake detection methods and evaluated their performance on our curated dataset to assess their suitability. The model weights of the methods were obtained from the respective authors. The detection performance results are summarized in TABLE~\ref{tab:summary3}, with a scatter-plot visualization presented in Fig.~\ref{fig:diagram_proposed_testing}(a).
\begin{table}[t!]
\renewcommand{\arraystretch}{1.2}  
  \centering
  \caption{Performance of deepfake detection methods for protection against Faceswap-GAN~\cite{fsg} and DiffSwap~\cite{zhao2023diffswap}.}
  \scalebox{1.0}
  {
  \begin{tabular}{lclcl}
\hline \hline
 Method (year)    & \ \ \ AUC & \ \ \ \ \ \ AUC\\ 
        & Faceswap-GAN~\cite{fsg}   & DiffSwap \cite{zhao2023diffswap}  \\ \hline
MesoNet (2018) \cite{afchar2018mesonet} & \ \ \ \  \textbf{0.937} & \ \ \ \ \ \ 0.522   \\
 DSP FWA (2019) \cite{li2018exposing} & \ \ \ \ 0.818 & \ \ \ \ \ \ 0.657 \\
 Capsule (2019) \cite{nguyen2019use} & \ \ \ \  0.624 & \ \ \ \ \ \ 0.653 \\
Xception (2019) \cite{chollet2017xception}  & \ \ \ \  0.792 & \ \ \ \ \ \ 0.641 \\
  Face X-ray (2020) \cite{li2020face}  & \ \ \ \  0.794 & \ \ \ \ \ \ \textbf{0.833} \\
    EfficientNet (2020) \cite{bonettini2020video}  & \ \ \ \  0.728 & \ \ \ \ \ \ 0.717 \\
    F\textsuperscript{3}-Net (2020) \cite{qian2020thinking}  & \ \ \ \  0.806 & \ \ \ \ \ \ 0.694 \\
    ICT-Ref (2022) \cite{dong2022protecting} & \ \ \ \  0.595 & \ \ \ \ \ \ 0.643 \\ 
    LSDA (2024) \cite{yan2024transcending}  & \ \ \ \  0.843 & \ \ \ \ \ \ 0.726 \\
        
 \hline 
  \end{tabular}
	}
  \label{tab:summary3}
\end{table}
Among the nine tested methods, we selected Xception~\cite{chollet2017xception} and EfficientNet~\cite{bonettini2020video} to evaluate the effectiveness of the double neural network operations for deepfake detection.
Although these two may not offer the best performance, they serve as powerful feature extractors for image-related tasks. In contrast, Face X-Ray, LSDA, and DSP-FWA are specialized feature extractors designed for detecting specific artifacts, such as blending boundaries or warping patterns. In addition, F\textsuperscript{3}-Net operates in the frequency domain rather than the image domain, whereas our double neural network-based method compares image features in the spatial domain. EfficientNet and Xception are well-known for their powerful spatial feature extraction capabilities across various tasks, making them ideal choices for this study.

The first baseline considered is the Xception~\cite{chollet2017xception} network trained on the FaceForensics++ dataset~\cite{rossler2019faceforensics++} with deepfake videos generated by four methods, including Faceswap~\cite{fs}. The second one is the EfficientNetAutoAttB4ST~\cite{bonettini2020video} network trained on the DFDC dataset~\cite{dolhansky2020deepfake}, a dataset consisting of deepfake videos generated by various popular face-swapping methods, such as Facewap-GAN~\cite{fsg}, StyleGAN~\cite{karras2019style}, Faceswap~\cite{fs}, and NTH~\cite{zakharov2019few}.

\subsection{Performance Gain}
We investigate the performance gains of a deepfake detector empowered with a double-deepfake operation. First, we evaluate the two baseline approaches on our test dataset without applying the double-deepfake technique against Faceswap-GAN~\cite{fsg} and DiffSwap~\cite{zhao2023diffswap}, as shown in TABLE~\ref{tab:summary} and TABLE~\ref{tab:summary2}.
\begin{table}[t!]
\renewcommand{\arraystretch}{1.2}  
  \centering
  \caption{Detection performance of the proposed and baseline methods against Faceswap-GAN~\cite{fsg} generated deepfakes.\vspace{0mm}}
  \scalebox{0.95}
  {
  \begin{tabular}{lclclcl}
\hline \hline
 Method    & \ \ \ AUC & \ AUC & AUC trimmed\\ 
        & Mean (SD)   & Median (IQR) &   \ Mean (10\%)   \\ \hline
 Xception \cite{chollet2017xception} &  0.792 (0.11) & 0.799 (0.14) & \ \ \ \ 0.799 \\
 Xception \cite{chollet2017xception} (tuned) &  0.887 (0.07) & 0.896 (0.09) & \ \ \ \ 0.894 \\
EfficientNet \cite{bonettini2020video} &  0.728 (0.13) & 0.733 (0.16) & \ \ \ \ 0.732   \\ 
 EfficientNet \cite{bonettini2020video} (tuned) &  0.920 (0.06) & 0.926 (0.07) & \ \ \ \ 0.927   \\ 
 \hline 
\textbf{Proposed} &  \textbf{0.940 (0.05)} & \textbf{0.958 (0.05)}  & \ \ \ \ \textbf{0.947} \\
\hline
\hline
  \end{tabular}
	}
  \label{tab:summary}
\end{table}
\begin{table}[t!]
\renewcommand{\arraystretch}{1.2}  
  \centering
  \caption{Detection performance of the proposed and baseline methods against DiffSwap~\cite{zhao2023diffswap} generated deepfakes.\vspace{0mm}}
  \scalebox{0.95}
  {
  \begin{tabular}{lclclcl}
\hline \hline
 Method    & \ \ \ AUC & \ AUC & AUC trimmed\\ 
        & Mean (SD)   & Median (IQR) &   \ Mean (10\%)   \\ \hline
 Xception \cite{chollet2017xception} &  0.641 (0.10) & 0.639 (0.17) & \ \ \ \ 0.641 \\
 Xception \cite{chollet2017xception} (tuned) &  1.000 (0.00) & 1.000 (0.00) & \ \ \ \ 1.000 \\
EfficientNet \cite{bonettini2020video} &  0.717 (0.10) & 0.724 (0.17) & \ \ \ \ 0.718   \\ 
 EfficientNet \cite{bonettini2020video} (tuned) &  1.000 (0.00) & 1.000 (0.00) & \ \ \ \ 1.000 \\
 \hline 
\textbf{Proposed} &  \textbf{1.000 (0.00)} & \textbf{1.000 (0.00)}  & \ \ \ \ \textbf{1.000} \\
\hline
\hline
  \end{tabular}
	}
  \label{tab:summary2}
\end{table}
To ensure a fair comparison with our proposed method, we conducted fine-tuning on these two baseline methods using our training dataset. This involved keeping the features frozen and training a classification layer on top of the features until the performance was saturated on the validation dataset. After finetuning, EfficientNetAutoAttB4ST~\cite{bonettini2020video} had an AUC mean of $0.920$ across identities with a sample standard deviation of $0.06$ against Faceswap-GAN.

\begin{figure}[!t]%
	\hspace{-2mm}
	{\includegraphics[width=1.01\linewidth,trim={0.1cm 0.1cm 0.49cm 0cm},clip]{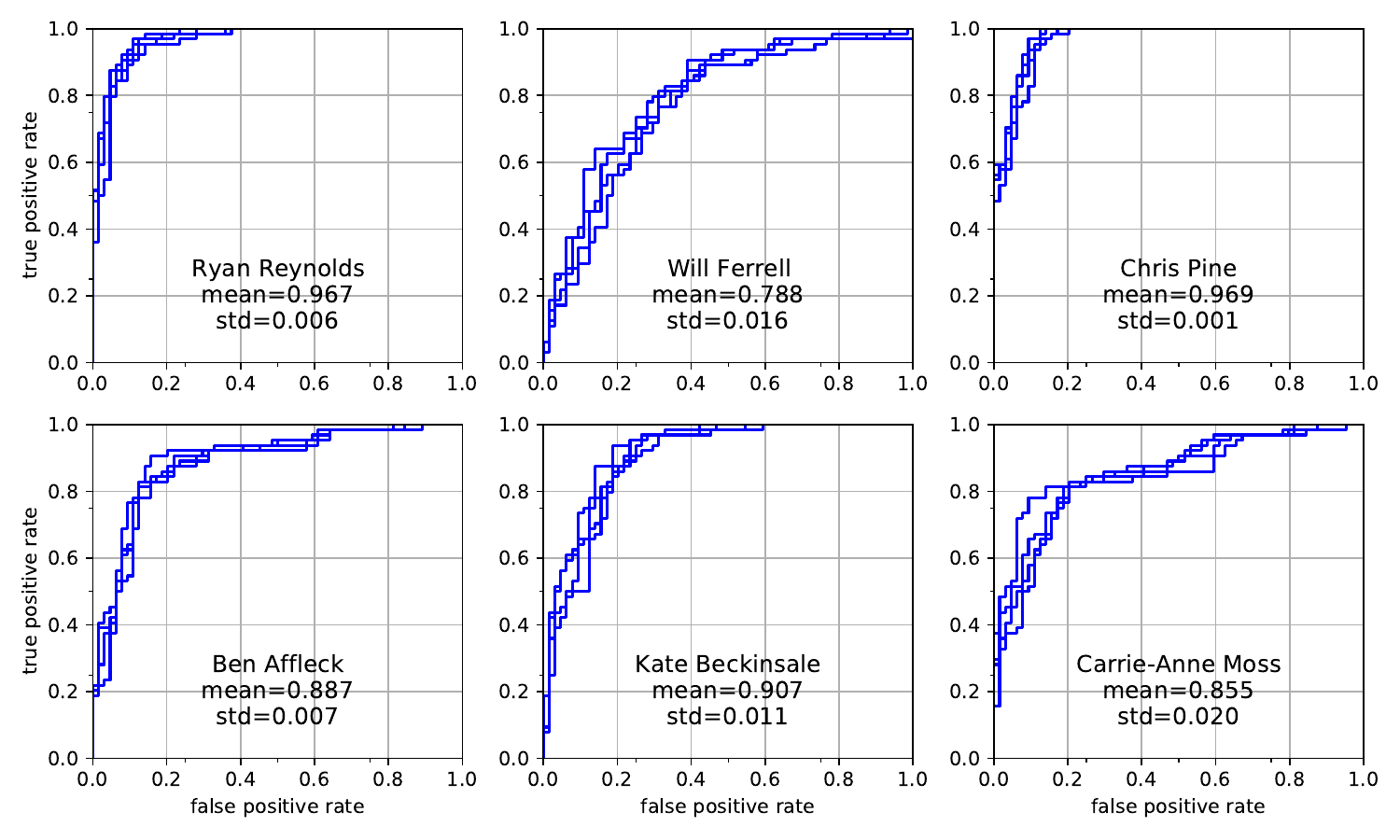}\vspace{-1mm}} 
    \caption{ROC curves for deepfake detection using the proposed method. Each plot contains results from a public figure, and each curve represents a trial of training the network. AUC values are large with small standard deviations, indicating good performance.}
	\vspace{-2mm}
    \label{fig:siamese_roc}
\end{figure}

We applied the double neural network operation to evaluate the idea of utilizing idempotency and identity conditioning and obtained the features from our trained identity-aware feature extractor. We concatenated those with the features of EfficientNetAutoAttB4ST~\cite{bonettini2020video}. TABLE~\ref{tab:summary} reveals that the proposed method can achieve an AUC mean of $0.940$ across identities, an increase of $0.020$ from Bonettini et al.~\cite{bonettini2020video}.
The AUC median across identities was $0.958$ with a gain of $0.032$ from the baseline~\cite{bonettini2020video}. The 10\%-trimmed mean was $0.947$ with a gain of $0.02$. The AUC standard deviation was reduced by $0.01$ or $17\%$, and the AUC interquartile range was reduced by $0.02$ or $29\%$ compared to the baseline~\cite{bonettini2020video}.
This result demonstrates that idempotency and identity conditioning can improve performance in validity and variation. The detection results on the test dataset for six of the $45$ public figures are shown in Fig.~\ref{fig:siamese_roc}.
The averaged AUC value among all public figures is $0.940$, and the sample standard deviation is $0.05$. We also performed $t$-tests, and the proposed method is significantly better (with $95\%$ confidence interval) than those of the off-the-shelf detectors in terms of AUC. The larger variance of the AUC values of the baseline methods implies that the deepfake detector may perform convincingly for one identity, but it has a greater risk of exhibiting unacceptable performance for others. This makes the baseline methods less attractive for journalism applications.

\subsection{Ablation Studies}
TABLE~\ref{tab:ablation_summary} displays the results of ablation studies.
In the first ablation study, we applied our idempotent strategy (with identity decoder) using the EfficientNetAutoAttB4ST features. In the second study, we concatenated the features from the identity-aware feature extractor with the features of EfficientNetAutoAttB4ST as we did in our proposed method and used a feedforward network to classify the images. The first ablation achieved the AUC mean of $0.926$, and the AUC median was $0.928$. The sample standard deviation and interquartile range were $0.05$ and $0.06$. The second ablation achieved the AUC mean of $0.893$, and the AUC median was $0.920$. The sample standard deviation and interquartile range were $0.10$ and $0.13$. The achieved AUC values are much lower compared to the proposed method. This confirms that the identity conditioning and idempotence strategy have synergy (positive interaction).
\begin{table}[t!]
\renewcommand{\arraystretch}{1.2}  
  \centering
    \caption{Ablation studies for the proposed method.\vspace{0mm}}
  \scalebox{1.0}
  {
  \begin{tabular}{lclclcl}
\hline \hline
 Method    & \ \ \ AUC & \ AUC & AUC trimmed\\ 
        & Mean (SD)   & Median (IQR) &   \ Mean (10\%)   \\ \hline
\textbf{Proposed} &  \textbf{0.940 (0.05)} & \textbf{0.958 (0.05)}  & \ \ \ \ \textbf{0.947} \\
Idempotence &  0.926 (0.05) & 0.928 (0.06)  & \ \ \ \ 0.932 \\ 
Identity-aware features &  0.893 (0.10) & 0.920 (0.13)  & \ \ \ \ 0.904 \\ 
\hline
\hline
  \end{tabular}
	}
  \label{tab:ablation_summary}
  \end{table}

\subsection{Experimental Verification of Near-Idempotence}
\label{sec:veri_near_idem}
Our proposed detection method leverages the near-idempotence property of the deepfake operator. Exact idempotence occurs when an altered image, passed through a deepfake generator, depicts no further changes. In the case of near-idempotence, the second operation would lead to small changes compared to the first operation. 
Let the residues be defined concerning raw data \( f \) as follows:
\begin{subequations}
\begin{align}
    e_0 &= R_{\text{recon}}(f) - f, \\
    e_1 &= R_{\text{recon}}(R_{\text{df}}(f)) - R_{\text{df}}(f).
\end{align}
\end{subequations}
To establish near-idempotence, we require \( \|e_0\|_2 \gg \|e_1\|_2 \) for all \( f \), \( R_{\text{df}} \), and \( R_{\text{recon}} \), where \( R_{\text{df}} \) represents the deepfake operation and \( R_{\text{recon}} \) represents the reconstruction operation.
In this subsection, we experimentally verify this property of deepfake generators. Specifically, we focus on two types of deepfake operations: Faceswap-GAN~(FG) and diffusion-based~(D) methods. Based on the choice and the order of deepfake operations applied to an image, we present our results within three categories as follows.

\subsubsection{\(R_{\mathrm{df}} = R_{\mathrm{recon}} = FG
\)}
\begin{figure}[!t]%
\centering
	{
\includegraphics[width=0.48\linewidth,trim={1.5cm 0.1cm 14.4cm 1.5cm},clip]{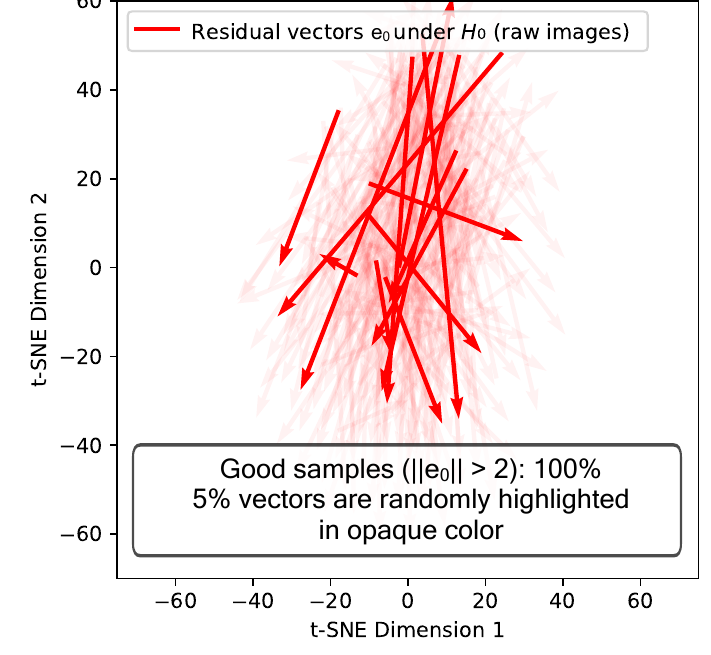}}\vspace{-1mm} 
    \vspace{1ex} 
    \makebox[0.55\textwidth][c]{\small(a)}\vspace{-3.4ex} 
    \vspace{0.3cm}
    \makebox[0.75\textwidth][c]{} 
\includegraphics[width=0.48\linewidth,trim={13.5cm 0.1cm 2.4cm 1.5cm},clip]{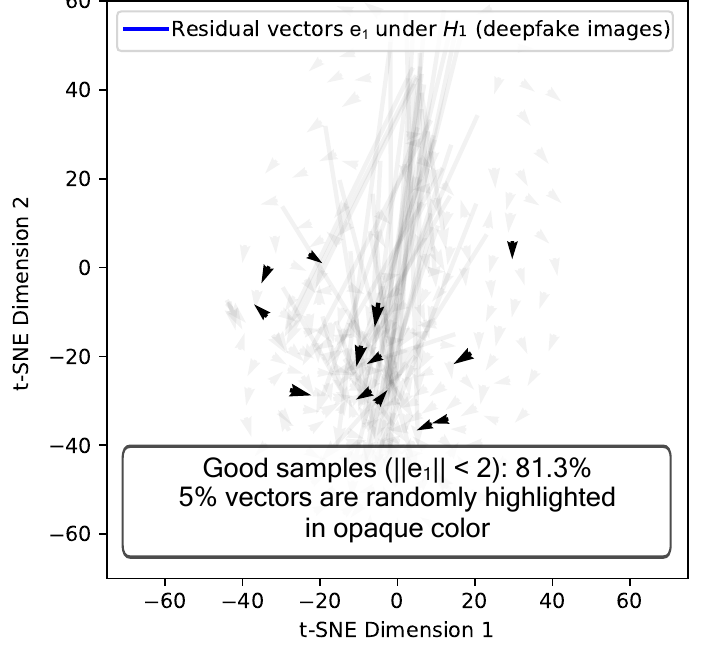}
\includegraphics[width=0.48\linewidth,trim={13.5cm 0.1cm 2.4cm 1.5cm},clip]{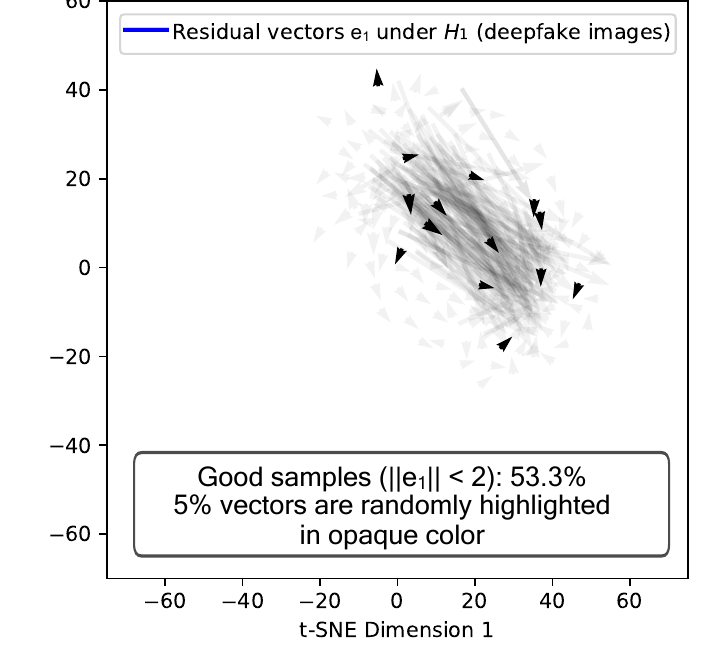}
    \vspace{1ex} 
    \makebox[0.55\textwidth][c]{\small(b)\hspace{4cm}(c)}\vspace{-3.4ex} 
    \makebox[0.75\textwidth][c]{} 
	\vspace{-2mm}
        \caption{t-SNE visualization for residual vectors (a)~\(e_0 = \mathrm{FG}(f) - f \) and  (b)~\(e_1 = \mathrm{FG}(\mathrm{FG}(f)) - \mathrm{FG}(f) \), when is Faceswap-GAN~(FG) is the deepfake operation in question. We randomly highlighted only 5\% vectors in opaque color for better viewing experiences. Contrasting (a) with (b), we note that residual vectors are shorter when the same operator is applied twice. The same can be concluded by contrasting residual vectors (a)~\(e_0 = \mathrm{FG}(f) - f \) and (c)~\(e_1 = \mathrm{FG}(\mathrm{D}(f)) - \mathrm{D}(f) \) when diffusion (D) is the deepfake operation in question.
        }
	\vspace{-0.5cm}
    \label{fig:vector_shifts_main}
\end{figure}

We trained a Faceswap-GAN face reconstructor for each identity using a subset of the available faces of that identity from the CelebDF dataset. Using these reconstructors for both deepfake generation and double neural network operation, we illustrate the residual vectors in Fig.~\ref{fig:vector_shifts_main}\,(a) and Fig.~\ref{fig:vector_shifts_main}\,(b). The residual vectors $e_0$ corresponding to an authentic image processed by an operator are shown in Fig.~\ref{fig:vector_shifts_main}\,(a). When the singly processed image is processed again by the same operator, the residual vectors $e_1$ are shown in Fig.~\ref{fig:vector_shifts_main}\,(b).  
These plots reveal that the first operation results in significant vector shifts, whereas the second operation leads to minimal shifts for $81.3\%$ of the samples. While Fig.~\ref{fig:vector_shifts_main} presents results for the CACD dataset, the corresponding plots for the CelebDF dataset are provided in the supplementary document.

\subsubsection{\(R_{\mathrm{df}} = R_{\mathrm{recon}} = {D}\)}
To verify the near-idempotence property of diffusion-based deepfake generators, we applied two consecutive diffusion operations to authentic images. In this case, the repeated operation yields small residual vectors for \(89.1\%\) of the samples. The residual vectors are demonstrated in the supplementary document.

\subsubsection{\(R_{\mathrm{df}} = {D} 
\,\,\&\,\, R_{\mathrm{recon}} = {FG}
\)}
In this case, the operations are diffusion followed by Faceswap-GAN. The residual vectors for the diffusion-generated deepfakes are shown in Fig.~\ref{fig:vector_shifts_main}\,(c). Here, the percentage of small residual vectors decreases to \(53.3\%\), indicating that applying a double neural network can be challenging when two different types of operations are involved. A careful design of the Siamese network is therefore necessary. Further analysis is provided in the supplementary document.
\section{Discussion}
\label{sec:discussion}
In this work, we have focused on two modern, popular, and off-the-shelf deepfake generation methods: Faceswap-GAN and DiffSwap. 
Our approach applies a deepfake operation specified by a forensic analyst and uses the norms of resulting residual vectors as a proxy to determine whether the deepfake operation is being applied for the first time or a second time. 
We examined scenarios where the forensic deepfake operation matches the original deepfake generation method and where the forensic operation differs.
Our results reveal that a Siamese detector trained under ideal conditions, where both operations are the same, is also effective when both operations switch to a new type. However, when the two operations differ from each other, the trained Siamese detector is less effective. This suggests the need for a more advanced Siamese detector capable of leveraging processing traces when the two operations are different.

Compared to end-to-end CNN-based classifiers, our proposed method targets deepfake detection for individuals, with main applications on public figures.
Although our method needs training the reconstruction models, the training can be done in advance for each public figure. 
For example, a journalist can train the reconstruction models for various candidates before they need to verify videos for reporting tasks.
Journalists may also share or collaboratively train detectors within their professional networks.
To let the detection system support a new individual, the journalist will need to train a reconstruction operator for that individual and then fine-tune the Siamese network.
\section{Conclusion and Future Work}\label{sec:conclusion}
In this work, we have proposed to use the method of double neural network operations and individual conditioning for deepfake detection. 
The proposed detector can achieve better detection performance than end-to-end CNN-based detectors on our curated dataset of public figures with identity labels. We have found that utilizing identity information can make the deepfake detector more reliable. 
We have also considered scenarios with mismatched first and second deepfake operations for real-world deepfake detection. Our results indicate that a Siamese detector trained on Faceswap-GAN is effective for diffusion-generated deepfake images, provided the additional deepfake operation is also diffusion-based. However, we identified a limitation of the proposed method when the forensic expert's deepfake operator differs, requiring the training of a new Siamese architecture for that specific combination. In future work, we aim to address this limitation by developing a generalized Siamese detector for deepfake detection.

\appendices

\bibliographystyle{IEEEtran}
\bibliography{main}

\clearpage

\begin{center}
\textbf{\Large Supplemental Material}    
\end{center}

\section{Advantage of Identity-Conditioned Feature Extraction} \label{sec:advantage_identity_aware}
\def\P{\operatorname{\mathbb{P}}}
Let us consider a set of images $S$ containing authentic and deepfake images. Each images is associated with an identity $k \in \{1,\dots,K\}$. $S$ may be decomposed into disjoint sets as follows:
\begin{equation}
S=\bigcup_{
k=1}^K S^{(k)} 
= S_{\text {auth}} \cup S_{\text {df }}
= S_0 \cup S_1,
\end{equation}
where $S^{(k)}$ is the set of all images belonging to individual $k$, 
$S_{\text{auth}}$ and $S_{\text{df}}$ are the sets of all authentic and deepfake images, respectively,
and $S_0$ and $S_1$ are the acceptance region and rejection region partitioned by a decision rule~\cite{van2004detection}.

\begin{figure*}[]
    \centering
    \resizebox{\textwidth}{!}{
    \begin{tabular}[b]{@{\hskip-5pt}c@{\hskip-5pt}@{\hskip-5pt}c@{\hskip-5pt}}
    \begin{tabular}{c}
    \smallskip
        \begin{subfigure}[t]{0.5\textwidth}
            \centering
            \includegraphics[width=\linewidth,trim={2cm 10cm 2cm 0cm},clip]{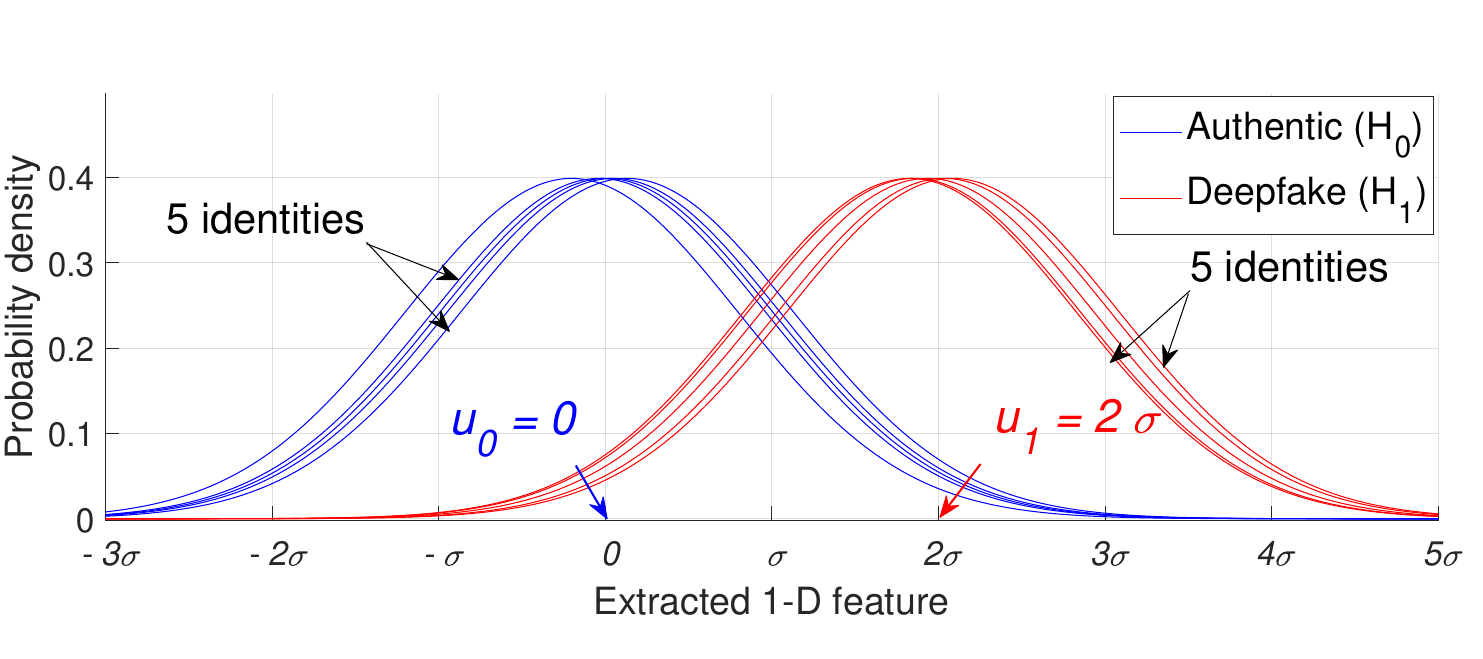}
            \caption{}
        \end{subfigure}\\
        \begin{subfigure}[t]{0.5\textwidth}
            \centering
            \includegraphics[width=\linewidth,trim={2cm 0cm 2cm 11cm},clip]{sec/figs/latex_fig_app_d_a_new.pdf}
            \caption{}
        \end{subfigure}
        
    \end{tabular}
&

\raisebox{-0.27cm}{
\begin{tabular}{c}
    \smallskip
    \begin{subfigure}{0.5\textwidth}
        \centering
        \smallskip
        \includegraphics[width=\linewidth,trim={1.5cm 0.1cm 1.85cm 0cm},clip]{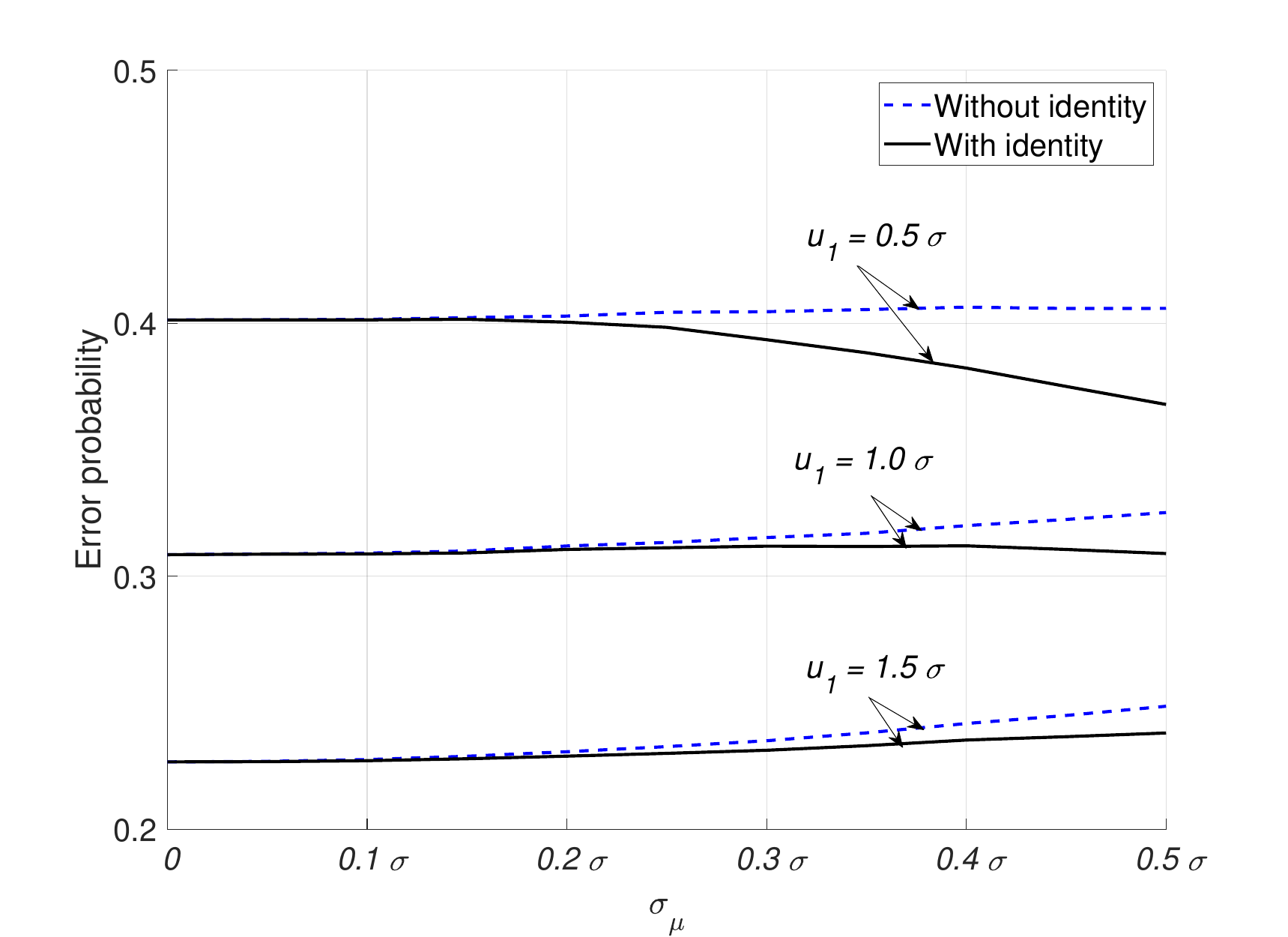}
        \caption{} 
    \end{subfigure}
\end{tabular}
}
\end{tabular}
}
\caption{Theory-driven simulation results: (a) probability density functions of extracted deepfake feature for $K = 5$ identities. Different identities' feature can have different distributions, as reflected by different $\mu_0^{(k)}$ with prior $u_0$ and different $\mu_1^{(k)}$ with prior $u_1$ for each identity $k$; (b) combined probability density function of extracted deepfake feature. If the identity information is not considered, then the individual distributions will mix into a single distribution; and (c) deepfake detection performance with and without the knowledge of the identity. Detection performance is better when the identity information is known. A larger gain can be achieved for the case of more unique individualized deepfake traces (larger $\sigma_\mu$) and more difficult detection problems (smaller $|u_1 - u_0|$).}
\label{fig:GMM_and_single}
\end{figure*}

Let us define $g: S \to \mathbb{R}$ as a powerful manifold-learning feature extractor for deepfake traces extraction so that the extracted 1-D feature $x = g(\text{f})$ for real images $\text{f} \in S_{\text{auth}}$ and fake images $\text{f} \in S_{\text{df}}$ exhibit different distributions. To facilitate our theoretical analysis and simulation, we consider the following hypotheses concerning an observation $x$ for individual $k$:
\begin{subequations}
\begin{align}
H_0:\ x = g\big(\text{f}\big) &\sim \mathcal{N}\big(\mu_0^{(k)}, \sigma^2\big), \quad \text{f} \in S^{(k)} \cap S_{\text {auth}}, \\ 
H_1:\ x = g\big(\text{f}\big) &\sim \mathcal{N}\big(\mu_1^{(k)}, \sigma^2\big),\quad  \text{f} \in S^{(k)} \cap S_{\text {df}},
\end{align}
\end{subequations}
where $\mu_0^{(k)}$ and $\mu_1^{(k)}$ have Gaussian priors, namely,
\begin{subequations}
\begin{align}
\mu_0^{(k)} &\sim \mathcal{N}\left(u_0, \sigma_\mu^2\right),\\
\mu_1^{(k)} &\sim \mathcal{N}\left(u_1, \sigma_\mu^2\right),
\end{align}
\end{subequations}
where we set $0 = u_0 < u_1 \in \mathbb{R}$ without loss of generality, and $\sigma_\mu^2$ is the variance of the priors. 
Fig.~\ref{fig:GMM_and_single}\,(a) illustrates the probability density functions~(PDFs) of $x = g(\text{f})$ under $H_0$ and $H_1$ for five individuals.
When identity information is unknown, the PDFs under each hypothesis merges into one as shown in Fig.~\ref{fig:GMM_and_single}\,(b).

The Bayes risk~\cite{van2004detection} for an arbitrary rejection region $S_1$ is defined as 
\begin{align}
r(S_1) = C_{10} \! \P(S_1 | H_0) \! \P(H_0)
\! + \! C_{01} \! \P(S_0 | H_1) \! \P(H_1),
\end{align}
where $\mathbb{P}(\cdot)$ is the probability measure, $C_{ij}$ is the cost incurred by choosing $H_i$ when $H_j$ is true, and $\P(H_i)$ is the prior. 
To focus on the effect of identity conditioning, we assume that the dataset $S$ is balanced, i.e., $\P(H_0) = \P(H_1) = 0.5$ and the incurred costs are the same, i.e., $C_{01} = C_{10} = 1$. With these assumptions, the Bayes risk is reduced to the overall error probability $P_\text{e}$.

We define $S_i^{(k)} = S_i \cap S^{(k)}$ to further segment the acceptance region $S_0$ and the rejection region $S_1$ by individuals:
\begin{subequations}
\begin{align}
P_\text{e} & = \frac{1}{2} \big[ \P(S_1 | H_0) 
+ \P(S_0 | H_1) \big] \\
& = \frac{1}{2} \Big[ \P \big( \cup_{k=1}^K S_1^{(k)} | H_0 \big) 
+ \P \big( \cup_{k=1}^K S_0^{(k)} | H_1 \big) \Big] \\
& = \frac{1}{2} \left[ \sum_{k=1}^K \P( S_1^{(k)} | H_0) + \sum_{k=1}^K \P( S_0^{(k)} | H_1) \right ] \label{eq:bayes-err} \\
& = \frac{1}{2} \bigg\{ 1 + \sum_{k=1}^K \left[ \P( S_1^{(k)} | H_0) - \P( S_1^{(k)} | H_1) \right] \bigg\}. \label{eq:expression-to-derive-LRT}
\end{align}
\end{subequations}
Standard hypothesis testing technique~\cite{van2004detection} allows us to derive from \eqref{eq:expression-to-derive-LRT} the optimal decision rule that minimizes the Bayes risk or error probability.
One can proceed with the derivation and the decision rule turns out to be separable for each individual $k$ and in the form of the likelihood ratio test, namely,
\begin{equation}
    S_1^{(k)} = \left\{ x > \frac{\mu_0^{(k)} + \mu_1^{(k)}}{2} = T^{(k)} \right\},
\end{equation}
where $T^{(k)}$ is the optimal decision threshold. 

Using the optimal decision rule, one can calculate the minimal error probability following ~\eqref{eq:bayes-err}:
\begin{subequations}
\begin{align}
P_\text{e}^{\text{ind}} 
= \frac{1}{2} \sum_{k=1}^K \Big[ &\P( S_1^{(k)} | H_0) + \P( S_0^{(k)} | H_1) \Big ] \\
= \frac{1}{2} \sum_{k=1}^K \Big[ 
& \P( S_1^{(k)} | S^{(k)}, H_0) \P(S^{(k)} | H_0) \nonumber\\
+ &\P( S_0^{(k)} | S^{(k)}, H_1) \P(S^{(k)} | H_1) \Big] \\
= \frac{1}{2K} \! \sum_{k=1}^K \! \Big[ \! 
& \P( S_1^{(k)} \! | S^{(k)}\!,\! H_0) \! + \! \P( S_0^{(k)} \! | S^{(k)}\!,\! H_1) \! \Big] \label{eq:complicated-condition-form}\\
= \frac{1}{2K} \sum_{k=1}^K \Big[ 
& 1 \! - \! \Phi\Big(\tfrac{T^{(k)}-\mu_0^{(k)}}{\sigma}\Big) \! + \! \Phi\Big( \tfrac{T^{(k)}-\mu_1^{(k)}}{\sigma}\Big) \Big] \\
= \frac{1}{K}\sum_{k=1}^K \,\, & \Phi\left( -d_k \right).   \quad \blacksquare
\label{eqn:indclosedform}
\end{align}
\end{subequations}
Here, \eqref{eq:complicated-condition-form} is due to the assumption that the identities are uniformly distributed over the dataset, i.e., $\P(S^{(k)} | H_0) = \mathbb{P}(S^{(k)} | H_1) = 1/K$,
$\Phi$ is the cumulative density function~(CDF) of standard Gaussian, and $d_k = \big(\mu_1^{(k)} - \mu_0^{(k)} \big) \big/ 2\sigma$.

In contrast, when there is no information about the identity, the hypothesis testing problem is reduced to the basic form as shown in Fig.~\ref{fig:GMM_and_single}\,(b). One can prove the following identity-agnostic optimal decision rule:
\begin{equation}
    S_1^{(k)} = \left\{ x > \frac{u_0 + u_1}{2} = T \right\},\ \forall k.
\end{equation}
The minimal error probability $P_\text{e}^{\text{com}}$ with all identities mixed is then given by:
\begin{subequations}
\begin{align}
P_\text{e}^{\text{com}} 
&= \frac{1}{2} \sum_{k=1}^K \Big[ \P( S_1^{(k)} | H_0) + \P( S_0^{(k)} | H_1) \Big ] \\
&= \frac{1}{2K} \sum_{k=1}^K \Big[ 
1 - \Phi\big(\tfrac{T-\mu_0^{(k)}}{\sigma}\big) + \Phi\big( \tfrac{T-\mu_1^{(k)}}{\sigma}\big) \Big].
\end{align}
\end{subequations}
Plugging in $T$ and 
using the second-order Taylor expansion on $\Phi(\cdot)$ around $d_k$, we obtain,
\begin{equation}
\begin{aligned}
P_\text{e}^{\text{com}} \approx P_\text{e}^{\text{ind}} + \frac{1}{2K} \sum_{k=1}^K \left[ -\Phi^{\prime \prime}\left(d_k\right) \right] \alpha_k^{2}. \quad \blacksquare
\label{eq:theoretical-result}
\end{aligned}
\end{equation}
Here, $\alpha_k = \big[(u_0-\mu_0^{(k)}) + (u_1-\mu_1^{(k)})\big] \big/ 2\sigma$, $\Phi^{\prime \prime}(\cdot)$ is the second-order derivative of $\Phi$, and $-\Phi^{\prime \prime}(d_k) >0$. This reveals that $P_\text{e}^{\text{com}}$ is larger (worse) than $P_\text{e}^{\text{ind}}$, highlighting the significance of identity conditioning for detection.

Fig.~\ref{fig:GMM_and_single}\,(c) demonstrates the result of $P_\text{e}^{\text{ind}}$ and $P_\text{e}^{\text{com}}$ generated by a large number of iterations for  $u_1-u_0 \in \{0.5\sigma,\, 1.0\sigma,\, 1.5\sigma\}$. It is observed that the performance is improved when the individual distributions are used by the detector and such effect is amplified with a larger $\sigma_{\mu}$ [i.e., more unique individualized deepfake traces; larger $|\alpha_k|$ as in \eqref{eq:theoretical-result}] and with a smaller $|u_1 - u_0|$ [i.e., more intrinsically difficult detection problems; smaller $d_k$ in \eqref{eq:theoretical-result} for $\Phi^{\prime \prime}(\cdot)$'s monotonically increasing interval on the positive half of the axis].
We used $K = 5$ identities for this simulation and verified via simulation that the performance is not sensitive to the choice of $K$.

\section{Fine-Grained Performance Analysis Over Identities}
\begin{figure*}[!t]%
    \centering
		\vspace{-0mm}
    \includegraphics[width=\linewidth,trim={1.90cm 0cm 2.8cm 0cm},clip]{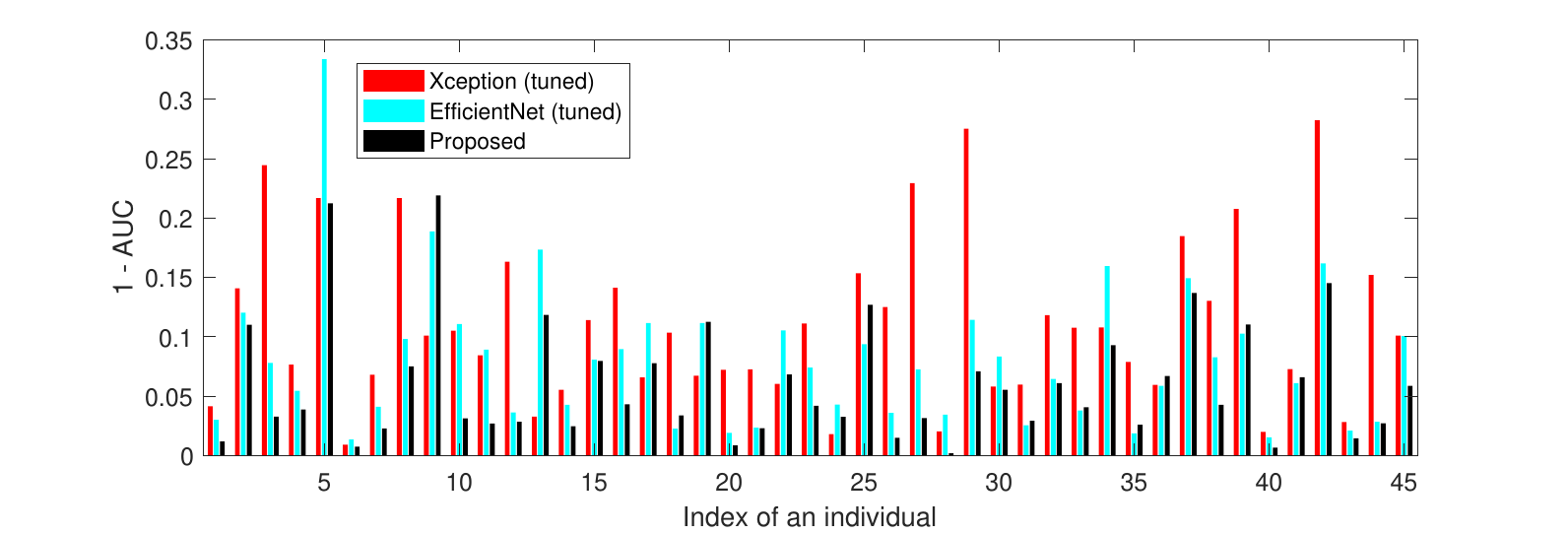}
    \vspace{-1mm}
    \caption{The performance of the deepfake detectors, measured in $1 - \text{AUC}$ (the smaller, the better), varies with the identities. The red and cyan peaks reveal that the baseline methods without utilizing identity information are less likely to perform well for specific individuals.}
	\vspace{-2mm}
    \label{fig:performance_variation}
\end{figure*}

The detection performance for an overall population of unknown composition may not be the most interesting metric from the perspective of a journalist when they target a specific celebrity or politician. Individualized deepfake detection proposed in this work allows more tailored optimization on an individual basis. The performance of the proposed individualized deepfake detector and two baseline methods for every public figure is shown in Fig.~\ref{fig:performance_variation}. The figure reveals that the performance of baseline methods is less consistent across the identity. For some identities, the performance of the baseline methods is significantly worse than their own average performance. This underscores the greater reliability and consistency of the proposed method in deepfake detection of public figures.

\section{Experimental Validation of Near-idempotence}
Our proposed detection method leverages the near-idempotence property of the deepfake operator. Exact idempotence occurs when an altered image, passed through a deepfake generator, depicts no further changes. In the case of near-idempotence, the second operation would lead to small changes compared to the first operation. 
Let the residues be defined concerning raw data \( f \) as follows:
\begin{subequations}
\begin{align}
    e_0 &= R_{\text{recon}}(f) - f, \\
    e_1 &= R_{\text{recon}}(R_{\text{df}}(f)) - R_{\text{df}}(f).
\end{align}
\end{subequations}
To establish near-idempotence, we require \( \|e_0\|_2 \gg \|e_1\|_2 \) for all \( f \), \( R_{\text{df}} \), and \( R_{\text{recon}} \), where \( R_{\text{df}} \) represents the deepfake operation and \( R_{\text{recon}} \) represents the reconstruction operation. In the upcoming two subsections, the first presents the detailed experimental results, while the second offers a short summary.

\subsection{Detailed Experimental Results}
In this subsection, we present detailed experimental results investigating the near-idempotence property of deepfake generators. Specifically, we focus on two types of deepfake operations: Faceswap-GAN~(FG) and diffusion-based~(D) methods. Based on the choice and the order of deepfake operations applied to an image, we present our results within three categories as follows.
\subsubsection{\(R_{\mathrm{df}} = R_{\mathrm{recon}} = \mathrm{FG}
\)}
\begin{figure*}[htbp]
    \centering
    \begin{subfigure}[b]{0.32\textwidth}
        \centering
        \includegraphics[width=\textwidth]{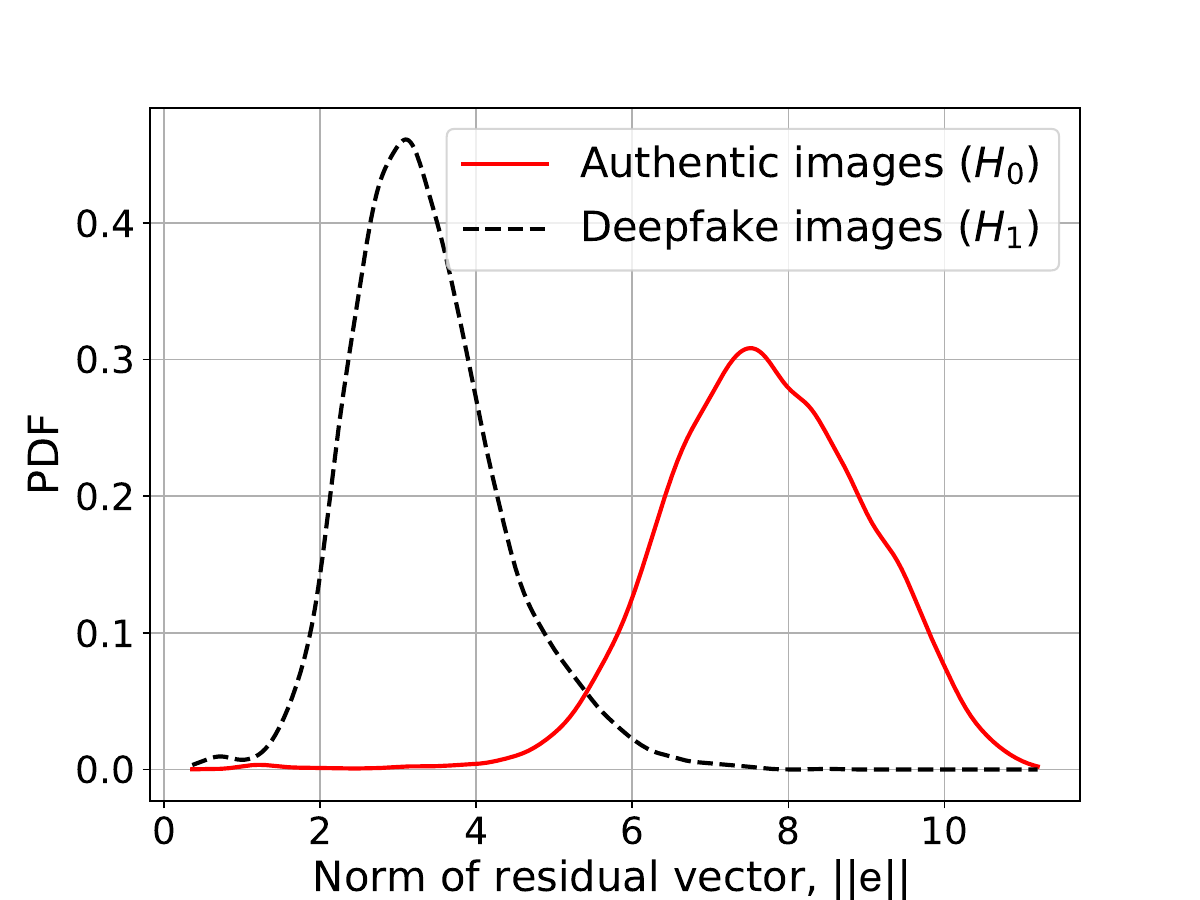}
        \label{fig:subfig1}
        \caption{\(\mathrm{FG}(f^{\mathrm{cdf}}),\mathrm{FG}(\mathrm{FG}(f^{\mathrm{cdf}}))\)}
    \end{subfigure}
    \begin{subfigure}[b]{0.33\textwidth}
        \centering
        \includegraphics[width=\textwidth]{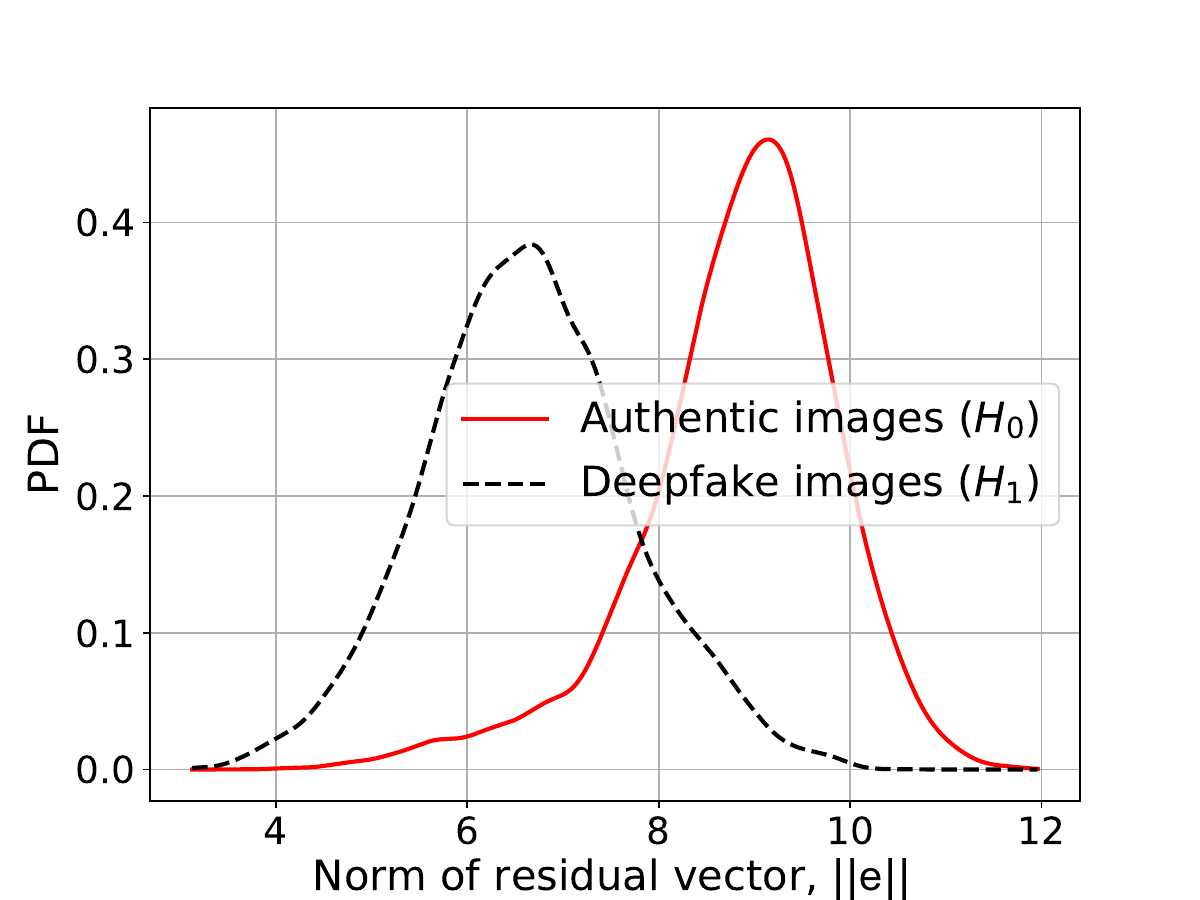}
        \label{fig:subfig2}
        \caption{\(\mathrm{D}(f^{\mathrm{cdf}}),\mathrm{D}(\mathrm{D}(f^{\mathrm{cdf}}))\)}
    \end{subfigure}
    \begin{subfigure}[b]{0.33\textwidth}
        \centering
        \includegraphics[width=\textwidth]{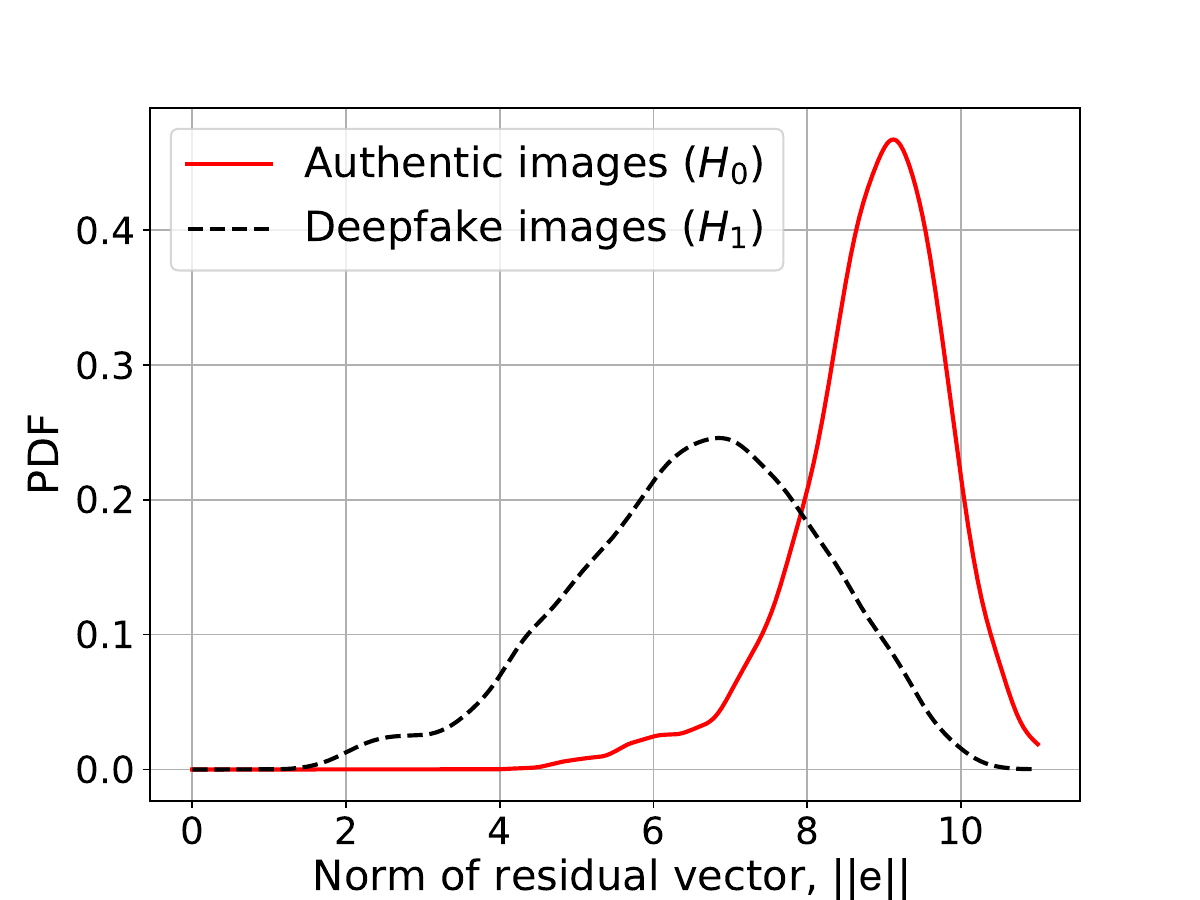}
        \label{fig:subfig3}
        \caption{\(\mathrm{FG}(f^{\mathrm{cdf}}),\mathrm{FG}(\mathrm{D}(f^{\mathrm{cdf}}))\)}
    \end{subfigure}

    \begin{subfigure}[b]{0.32\textwidth}
        \centering
        \includegraphics[width=\textwidth]{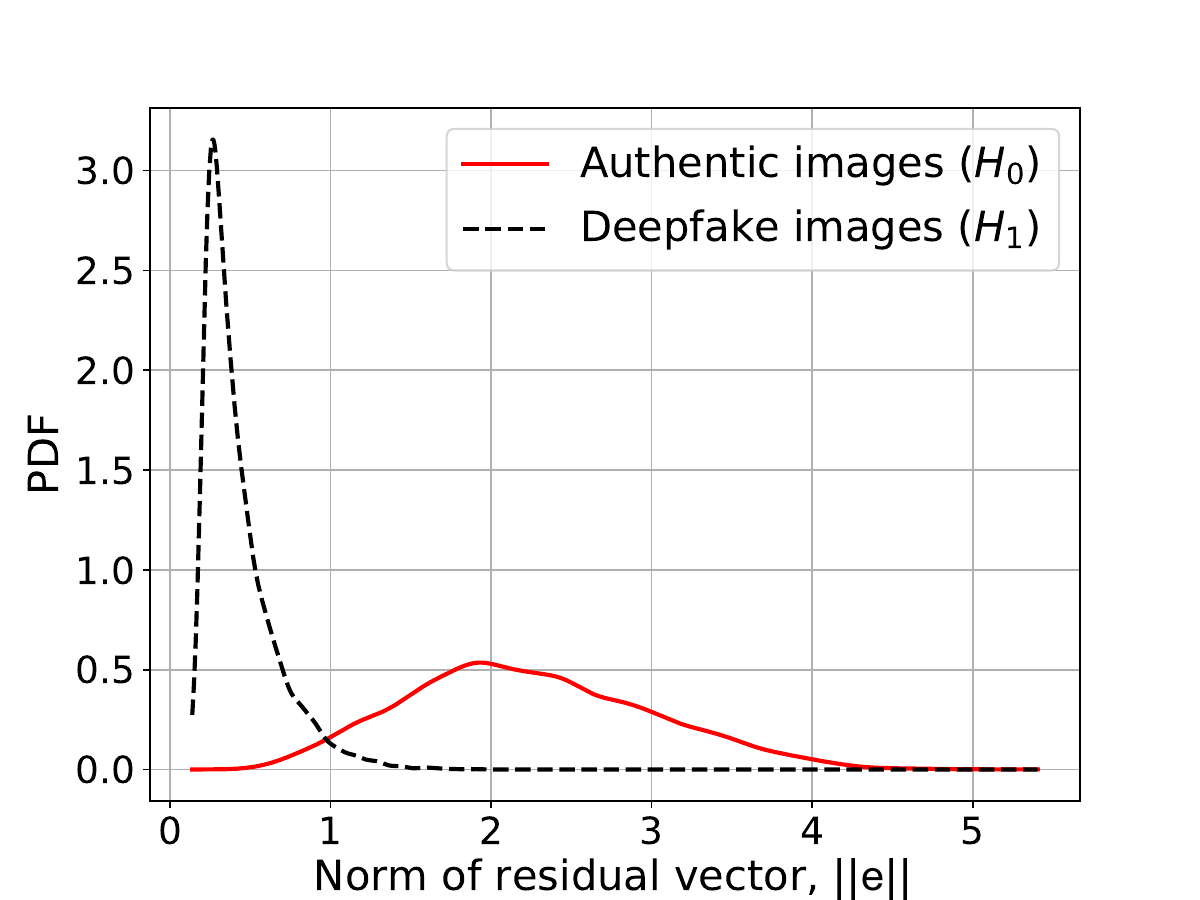}
        \label{fig:subfig1}
        \caption{\(S(\mathrm{FG}(f^{\mathrm{cdf}})),S(\mathrm{FG}(\mathrm{FG}(f^{\mathrm{cdf}})))\)}
    \end{subfigure}
    \begin{subfigure}[b]{0.33\textwidth}
        \centering
        \includegraphics[width=\textwidth]{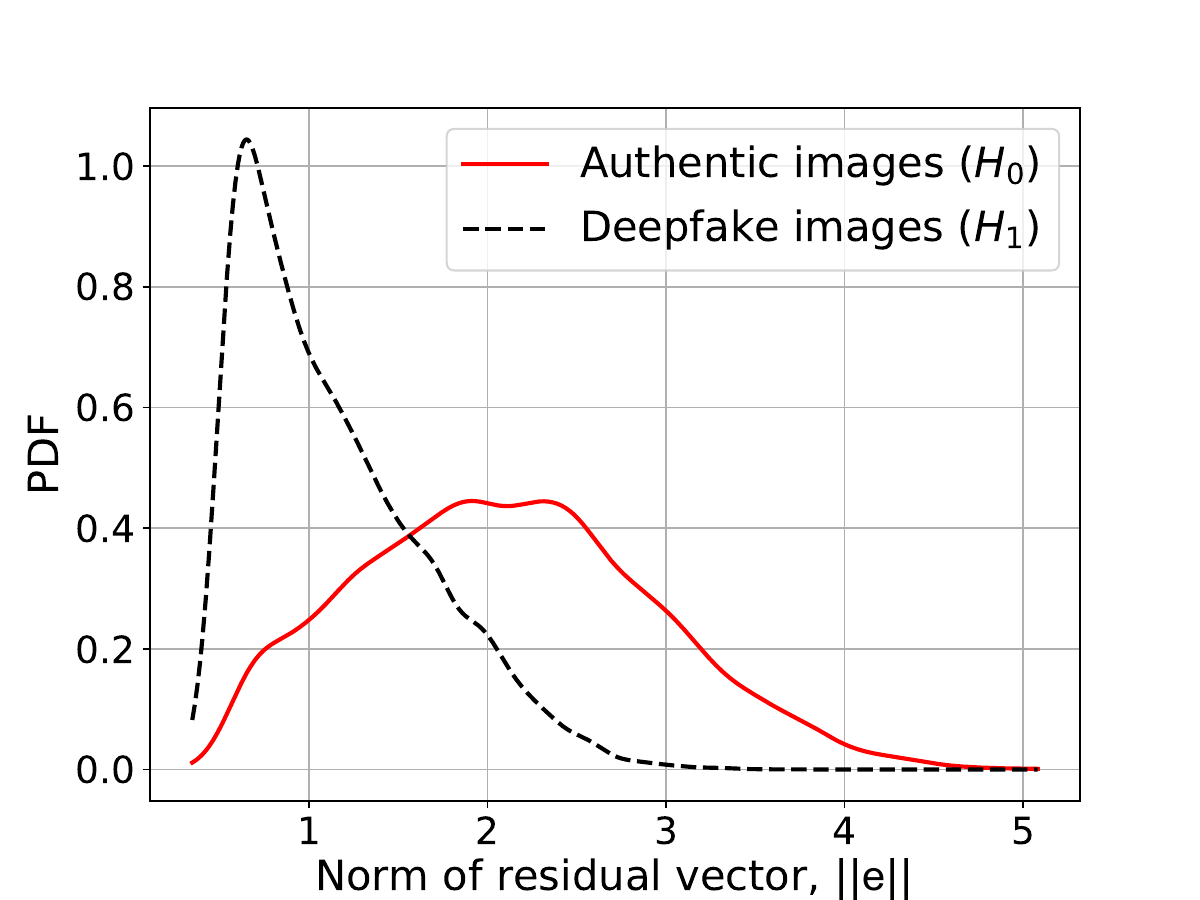}
        \label{fig:subfig2}
        \caption{\(S(\mathrm{D}(f^{\mathrm{cdf}})),S(\mathrm{D}(\mathrm{D}(f^{\mathrm{cdf}})))\)}
    \end{subfigure}
    \begin{subfigure}[b]{0.33\textwidth}
        \centering
        \includegraphics[width=\textwidth]{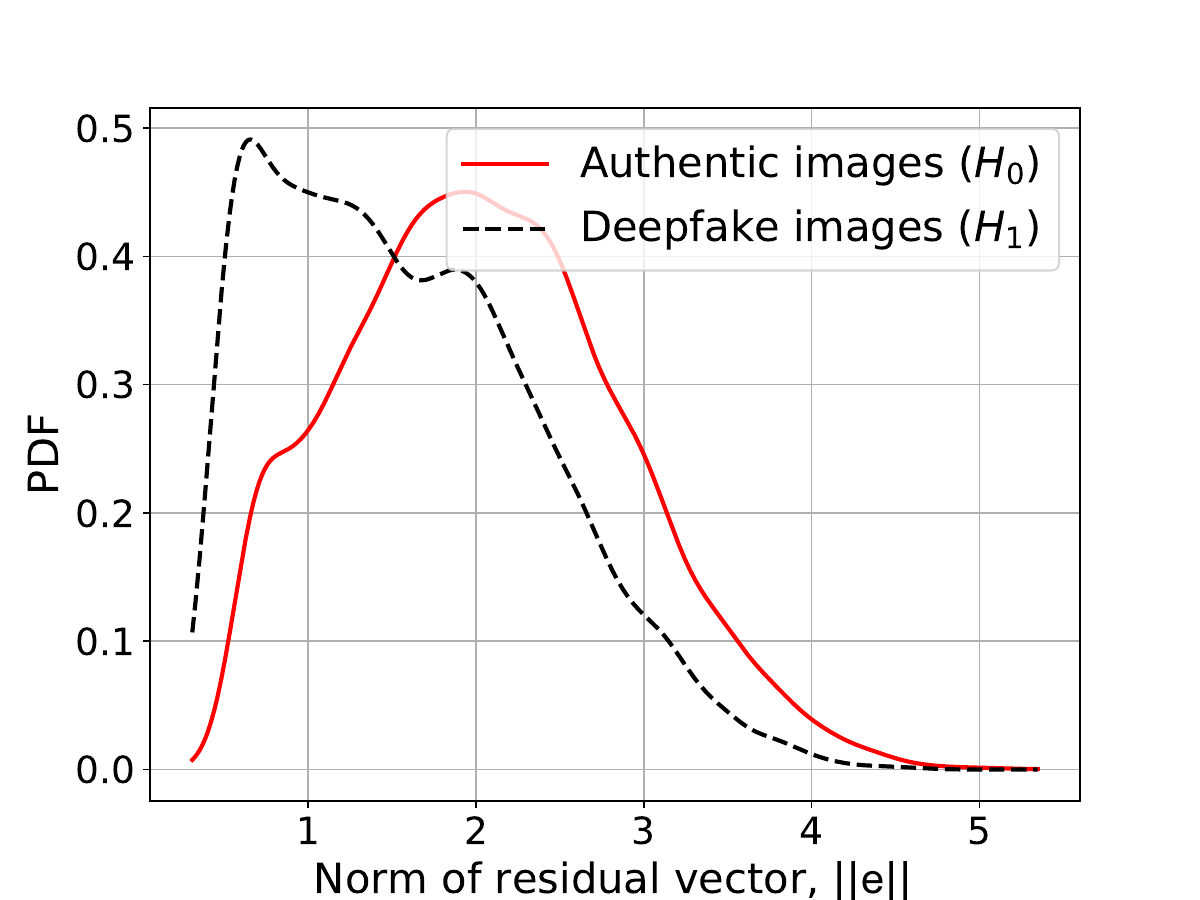}
        \label{fig:subfig3}
        \caption{\(S(\mathrm{FG}(f^{\mathrm{cdf}})),S(\mathrm{FG}(\mathrm{D}(f^{\mathrm{cdf}})))\)}
    \end{subfigure}

    \begin{subfigure}[b]{0.32\textwidth}
        \centering
        \includegraphics[width=\textwidth]{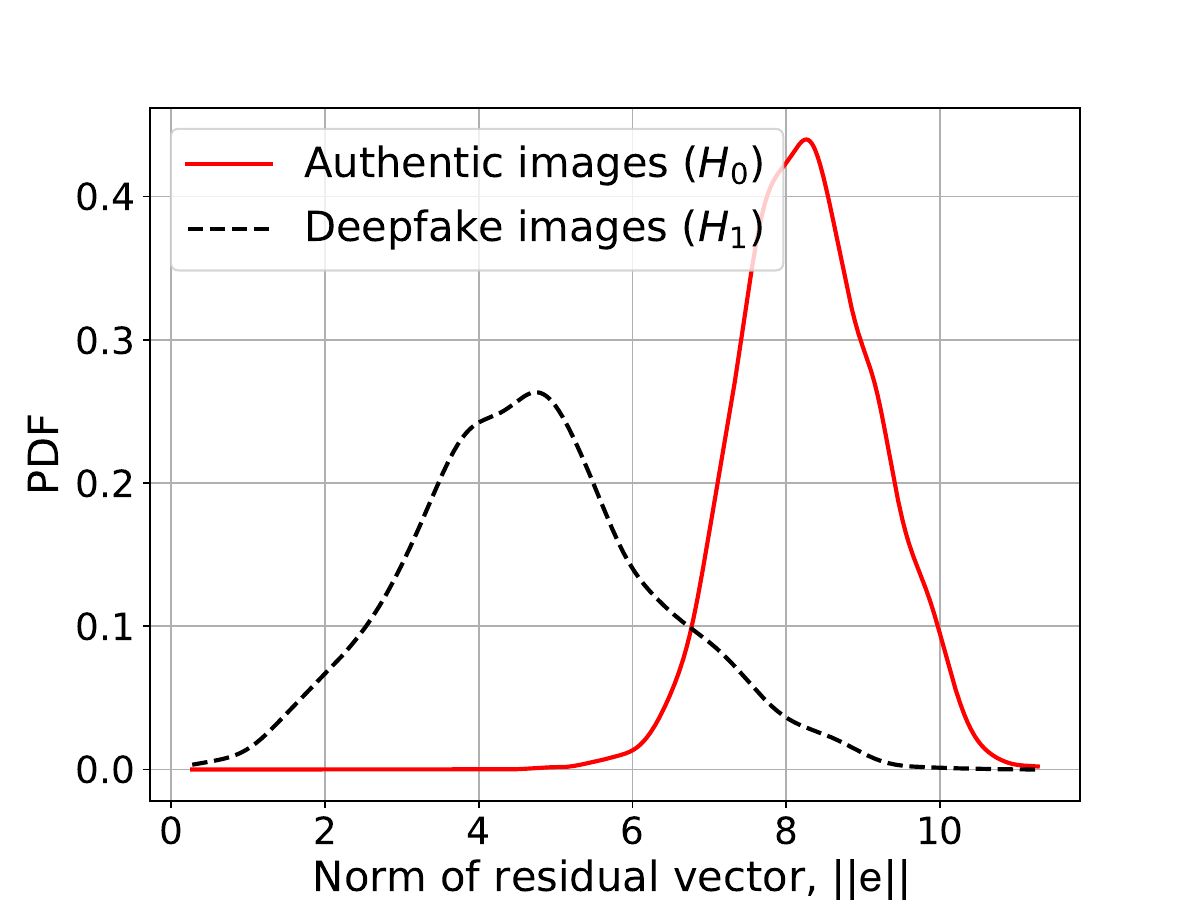}
        \label{fig:subfig1}
        \caption{\(\mathrm{FG}(f^{\mathrm{cacd}}),\mathrm{FG}(\mathrm{FG}(f^{\mathrm{cacd}}))\)}
    \end{subfigure}
    \begin{subfigure}[b]{0.33\textwidth}
        \centering
        \includegraphics[width=\textwidth]{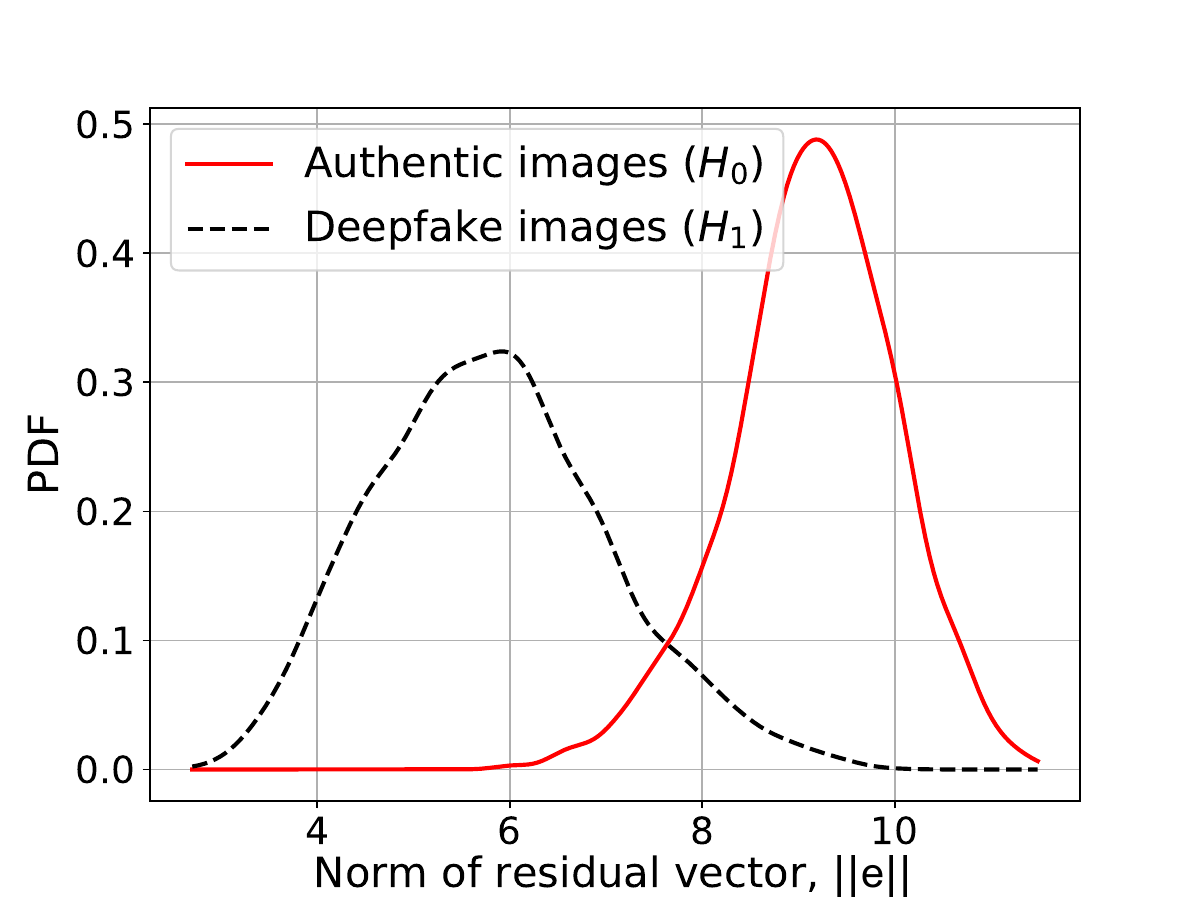}
        \label{fig:subfig2}
        \caption{\(\mathrm{D}(f^{\mathrm{cacd}}),\mathrm{D}(\mathrm{D}(f^{\mathrm{cacd}}))\)}
    \end{subfigure}
    \begin{subfigure}[b]{0.33\textwidth}
        \centering
        \includegraphics[width=\textwidth]{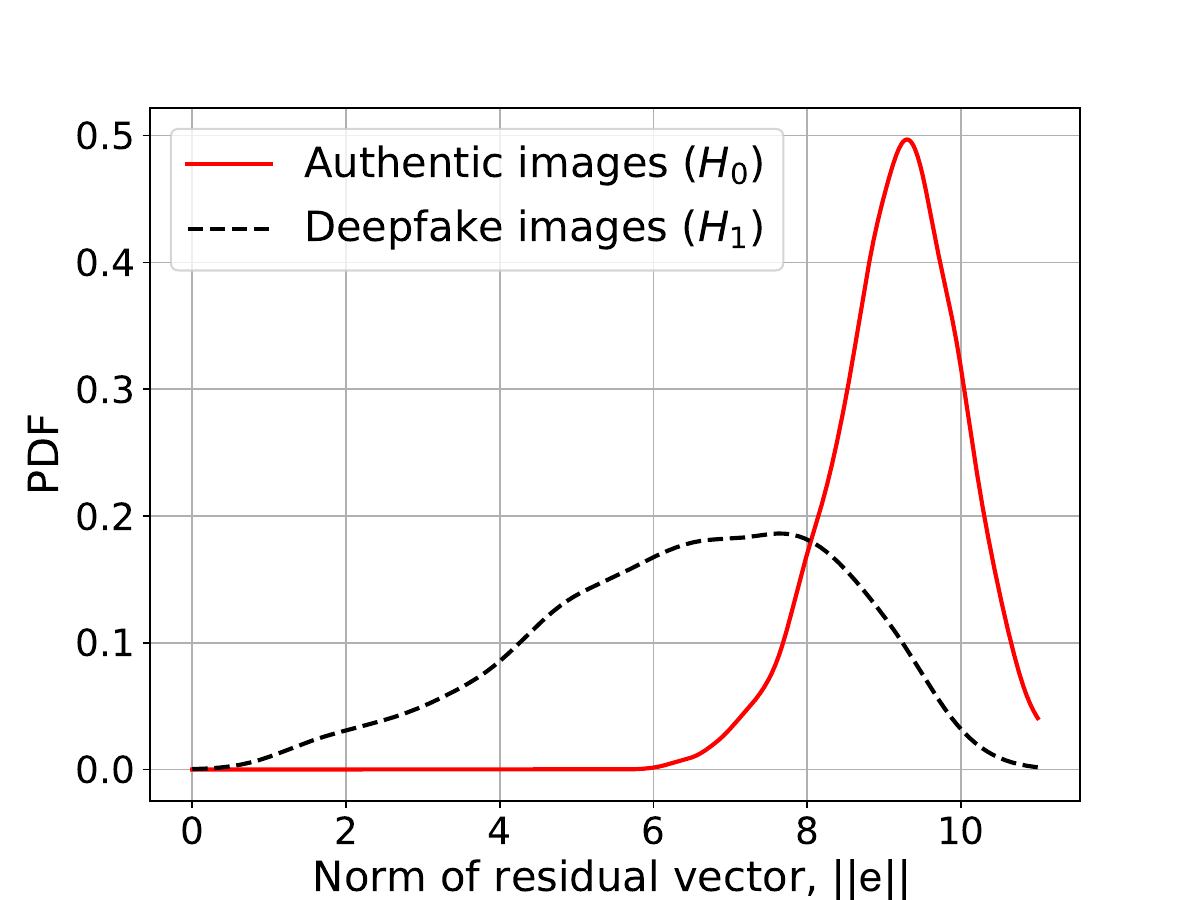}
        \label{fig:subfig3}
        \caption{\(\mathrm{FG}(f^{\mathrm{cacd}}),\mathrm{FG}(\mathrm{D}(f^{\mathrm{cacd}}))\)}
    \end{subfigure}

    \begin{subfigure}[b]{0.32\textwidth}
        \centering
        \includegraphics[width=\textwidth]{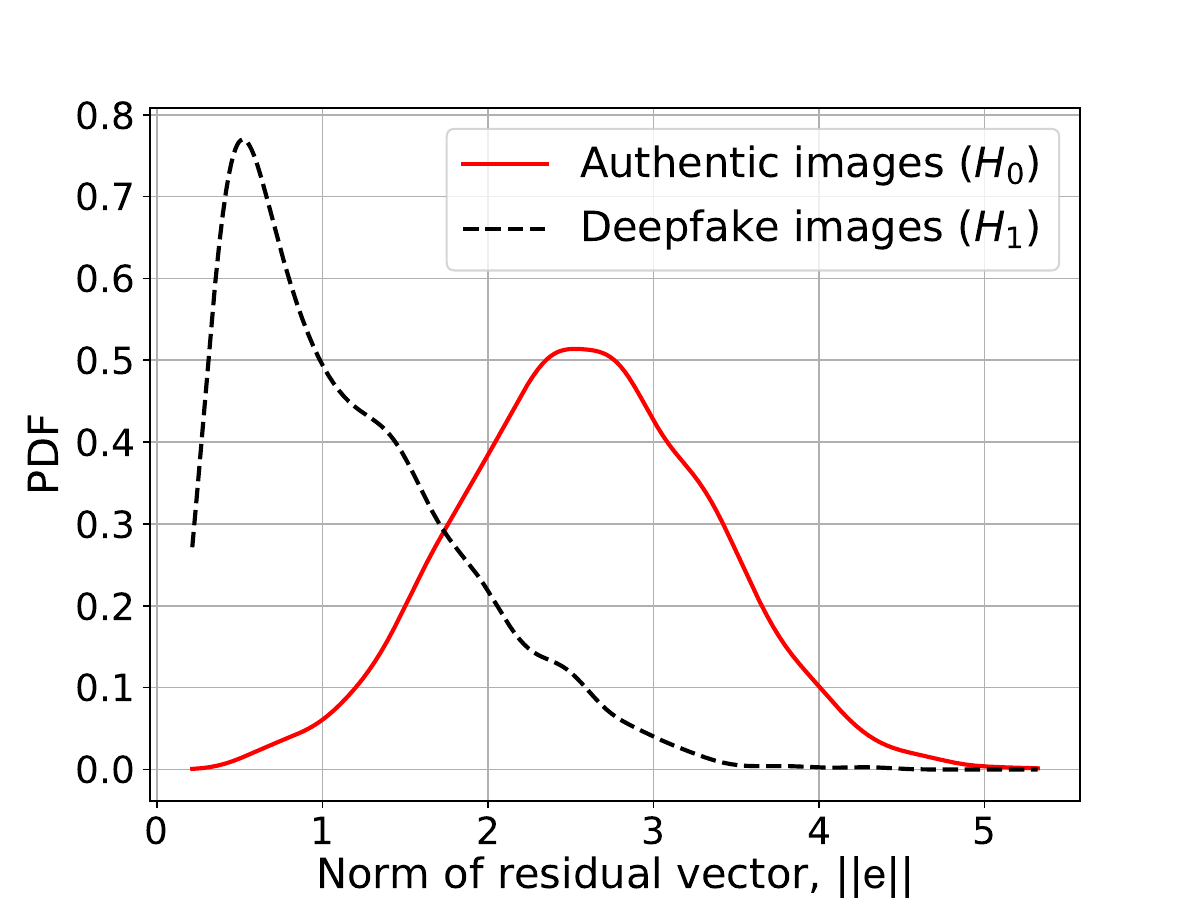}
        \label{fig:subfig1}
        \caption{\(S(\mathrm{FG}(f^{\mathrm{cacd}})),S(\mathrm{FG}(\mathrm{FG}(f^{\mathrm{cacd}})))\)}
    \end{subfigure}
    \begin{subfigure}[b]{0.33\textwidth}
        \centering
        \includegraphics[width=\textwidth]{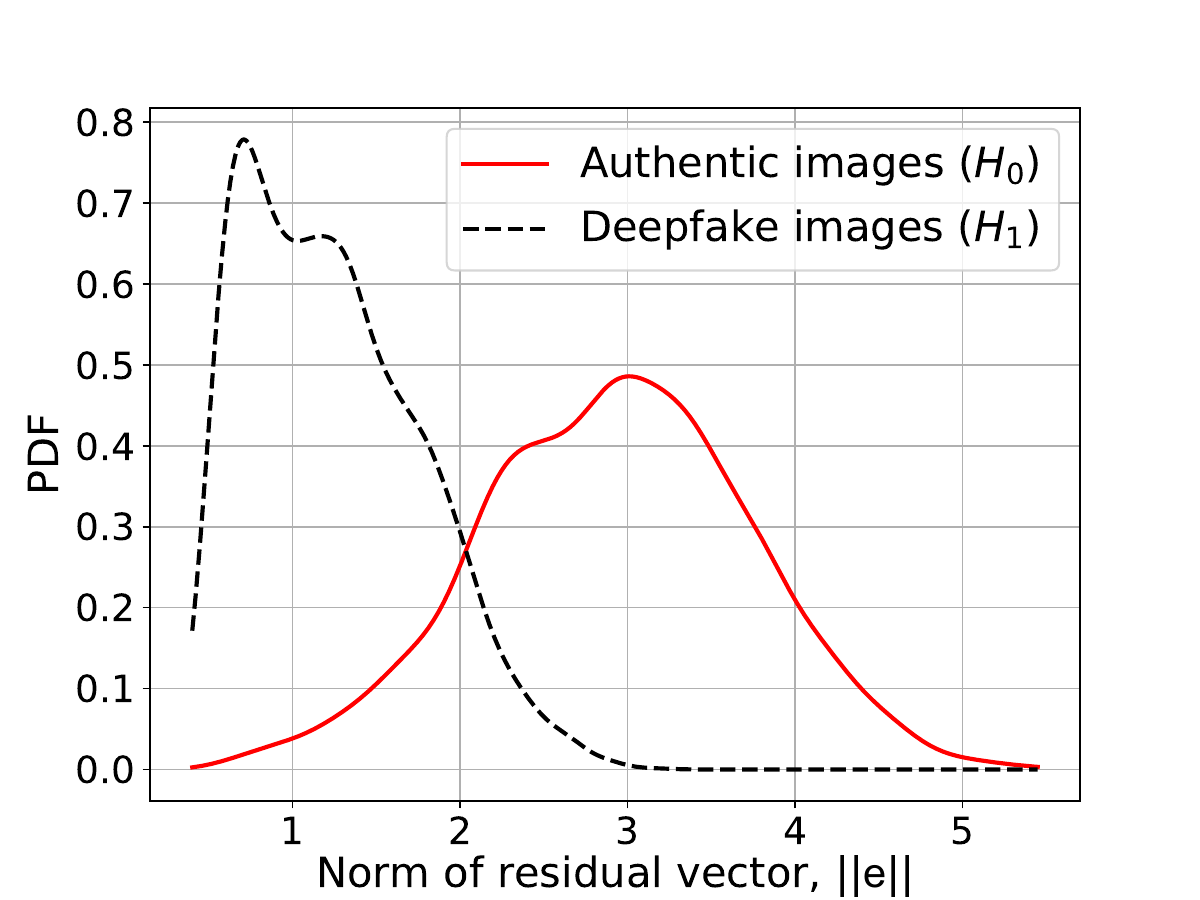}
        \label{fig:subfig2}
        \caption{\(S(\mathrm{D}(f^{\mathrm{cacd}})),S(\mathrm{D}(\mathrm{D}(f^{\mathrm{cacd}})))\)}
    \end{subfigure}
    \begin{subfigure}[b]{0.33\textwidth}
        \centering
        \includegraphics[width=\textwidth]{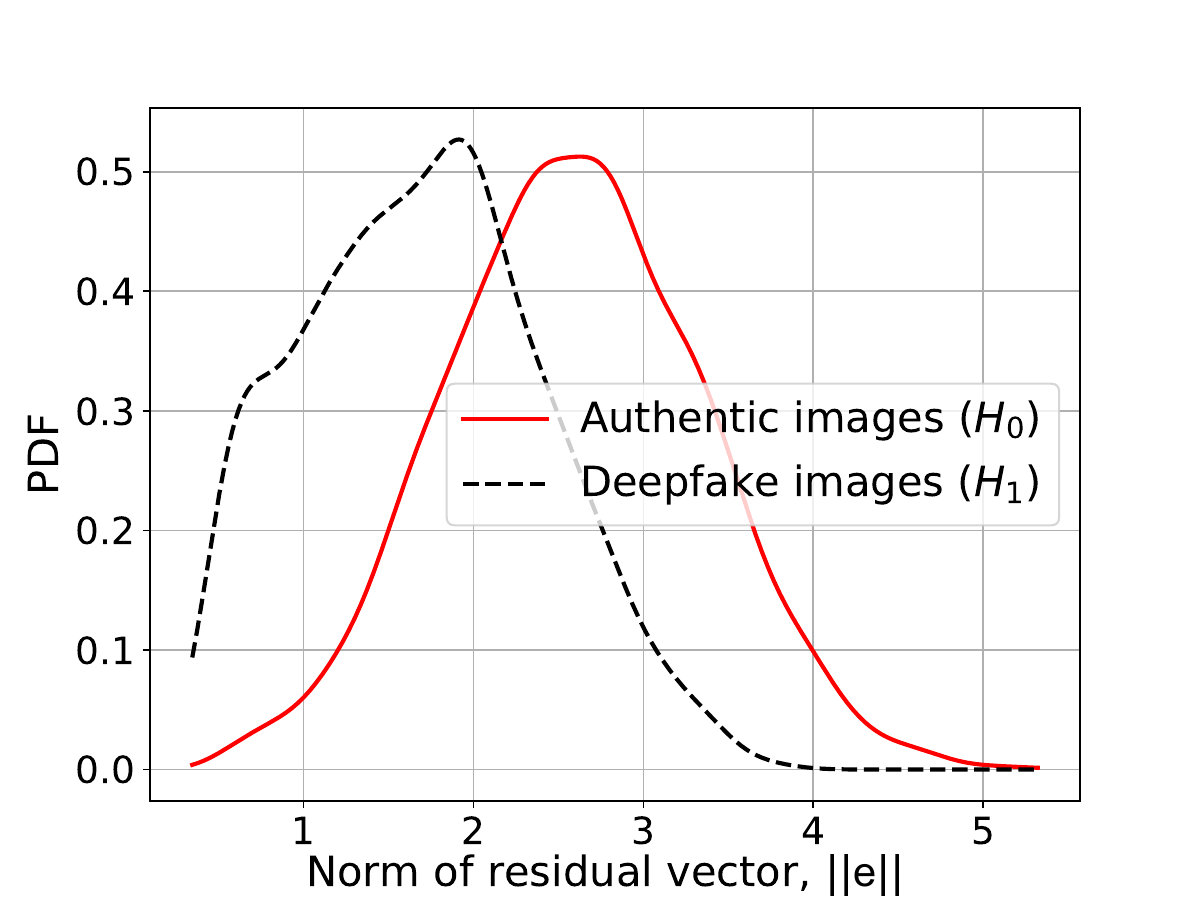}
        \label{fig:subfig3}
        \caption{\(S(\mathrm{FG}(f^{\mathrm{cacd}})),S(\mathrm{FG}(\mathrm{D}(f^{\mathrm{cacd}})))\)}
    \end{subfigure}
    \caption{PDF plots of residual vector norms. Each subplot is labeled as ``Result of first operation, Result of second operation," where \(\mathrm{CDF}, \mathrm{CACD}, \mathrm{FG}, \mathrm{D}, S\) denote CelebDF, CACD, Faceswap-GAN, Diffusion, and Siamese network, respectively.}
    \label{fig:mainfigure_pdf}
\end{figure*}

We learned a Faceswap-GAN face reconstructor for each of the identities from a subset of the available faces of that identity in the CelebDF dataset. In TABLE~\ref{tab:idempotence_summary}, we present the CDF values of the norm of the feature vector residuals at three threshold points for the first and second operations.
\begin{table*}[!b]
\renewcommand{\arraystretch}{1.2}  
\centering
\caption{Vector Shifting Statistics for two Faceswap-GAN Operations.\vspace{-3mm}}
\vspace{2mm}
\scalebox{1.0}
{
\begin{tabular}{|c|cc|cc|}
\hline
\multirow{2}{*}{Dataset} & \multicolumn{2}{c|}{\begin{tabular}[c]{@{}c@{}}Norm of Feature vector residual \(\|e\|\)\\ \textless 5.5, \textless 6.0, \textless 6.5\end{tabular}} & \multicolumn{2}{c|}{\begin{tabular}[c]{@{}c@{}}Norm of Siamese Feature Vector Residual \(\|e\|\)\\ \textless 1.25, \textless 1.5, \textless 1.75\end{tabular}} \\ \cline{2-5} 
                         & \multicolumn{1}{c|}{First Operation \(\|e_0\|\)}                                         & Second Operation \(\|e_1\|\)                                         & \multicolumn{1}{c|}{First Operation \(\|e_0\|\)}                                              & Second Operation \(\|e_1\|\)                                              \\ \hline
CelebDF                  & \multicolumn{1}{c|}{3.74 \%, 8.01 \%, 16.81 \%}                              & 97.12 \%, 98.80 \%, 99.50 \%                             & \multicolumn{1}{c|}{10.77 \%, 18.59 \%, 29.36 \%}                                   & 98.78 \%, 99.64 \%, 99.92 \%                                  \\ \hline
CACD                     & \multicolumn{1}{c|}{0.03 \%, 0.35 \%, 1.94 \%}                               & 72.29 \%, 79.86 \%, 86.11 \%                             & \multicolumn{1}{c|}{3.82 \%, 7.36\%, 13.57 \%}                                   & 62.33 \%, 73.19 \%, 81.18 \%                                  \\ \hline
\end{tabular}
}
\label{tab:idempotence_summary}
\end{table*}
For near-idempotence-based deepfake image detection to be effective, it is essential to observe a significant difference in the CDF values between the first and second operations. The CDF values for the second operation should be close to $1$ for validating the near-idempotence property. Our results show that this is true for both CelebDF and CACD datasets, with CelebDF demonstrating superior performance. The PDF plots of feature vector residual norms for two Faceswap-GAN operations are shown in the four subfigures of the first column in Fig.~\ref{fig:mainfigure_pdf}.
From TABLE~\ref{tab:idempotence_summary} and Fig.~\ref{fig:mainfigure_pdf}, we also observe that the feature Vector residual is biased based on the dataset. Consequently, slightly higher threshold values could be chosen for the CACD dataset to better separate the first and second operations. Since the deepfake operator is trained on a portion of the CelebDF dataset, the deepfake images generated using the CelebDF dataset are expected to be more realistic and, hence, dangerous. However, the separation method on vector residual also works better on the CelebDF dataset, mitigating its vulnerability. TABLE~\ref{tab:idempotence_summary} also demonstrates that the Siamese feature vectors with a reduced length of only $50$ can effectively capture the separation capability compared to the original feature vector of length $1856$.

\begin{figure}[!t]%
    \begin{subfigure}[b]{\linewidth}
        \centering
        \includegraphics[width=1.01\linewidth,trim={0.3cm 0.2cm 1.3cm 1.5cm},clip]{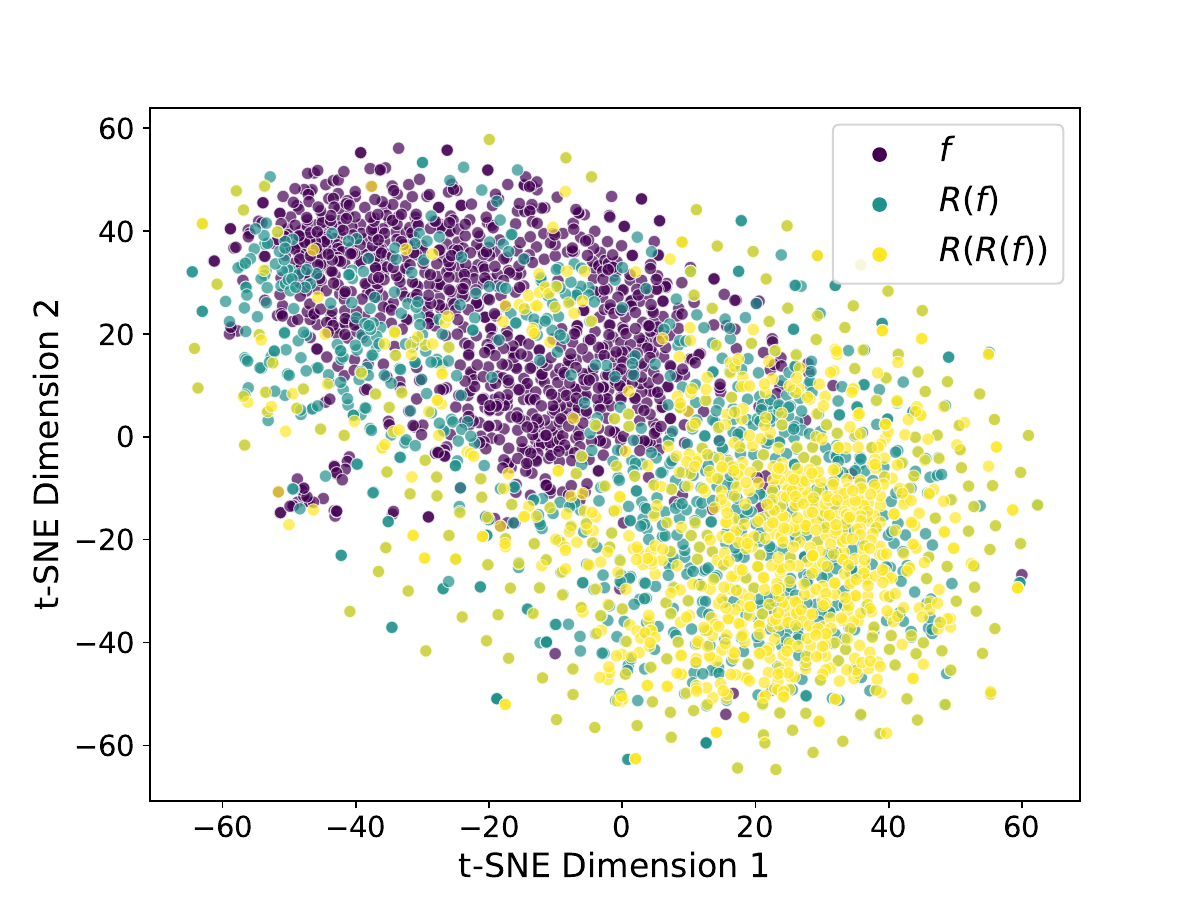}
        \caption{}
        \label{fig_intro2:image1}
        \vspace{0.3cm}
    \end{subfigure}
	\hspace{-2mm}
	{\includegraphics[width=1.01\linewidth,trim={1.5cm 0.1cm 2.4cm 1.5cm},clip]{sec/figs/vector_shift_cacd3.pdf}}\vspace{-1mm} 
    \vspace{1ex} 
    \makebox[0.28\textwidth][c]{(b)}\vspace{-3.4ex} 
    \makebox[0.75\textwidth][c]{(c)} 
        \caption{(a) t-SNE visualization of the feature space of authentic faces from CACD \(f\), faceswapped images generated by FaceSwap-GAN \(\mathrm{FG}(f)\), and the doubly processed image generated by the learned reconstruction operator \(\mathrm{FG}(\mathrm{FG}(f))\). The sample densities of the first and the second classes are visually separable, while the second and third classes exhibit considerable overlap. (b)~\(e_0 = \mathrm{FG}(f) - f \) in the t-SNE domain, (c)~\(e_1 = \mathrm{FG}(\mathrm{FG}(f)) - \mathrm{FG}(f) \) in the t-SNE domain. Vectors in (b) are significantly larger compared to those in (c).}
	\vspace{-2mm}
    \label{fig:vector_shifts_cacd}
\end{figure}

\begin{figure}[!t]%
    \begin{subfigure}[b]{\linewidth}
        \centering
        \includegraphics[width=1.01\linewidth,trim={0.3cm 0.2cm 1.3cm 1.5cm},clip]{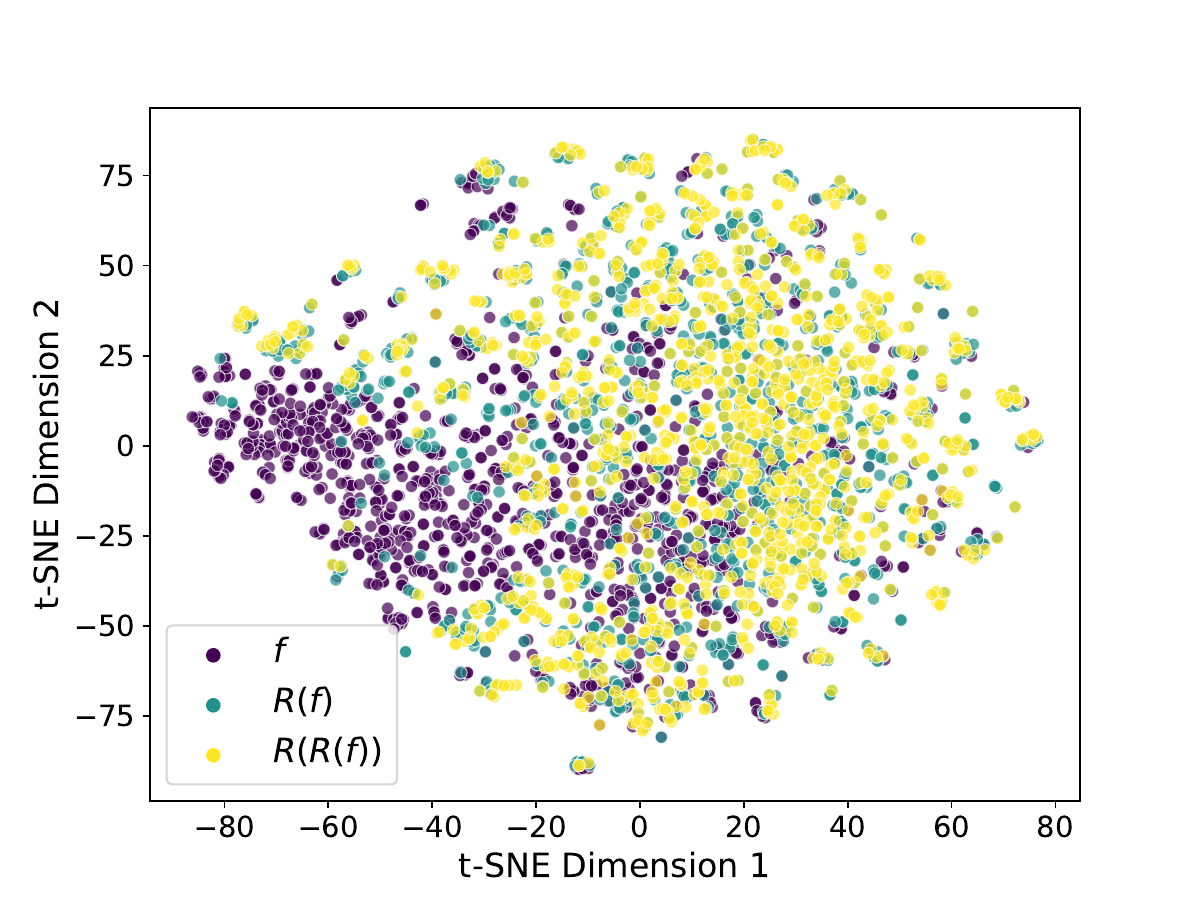}
        \caption{}
        \label{fig_intro2:image1}
        \vspace{0.3cm}
    \end{subfigure}
	\hspace{-2mm}
	{\includegraphics[width=1.01\linewidth,trim={1.5cm 0.1cm 2.4cm 1.5cm},clip]{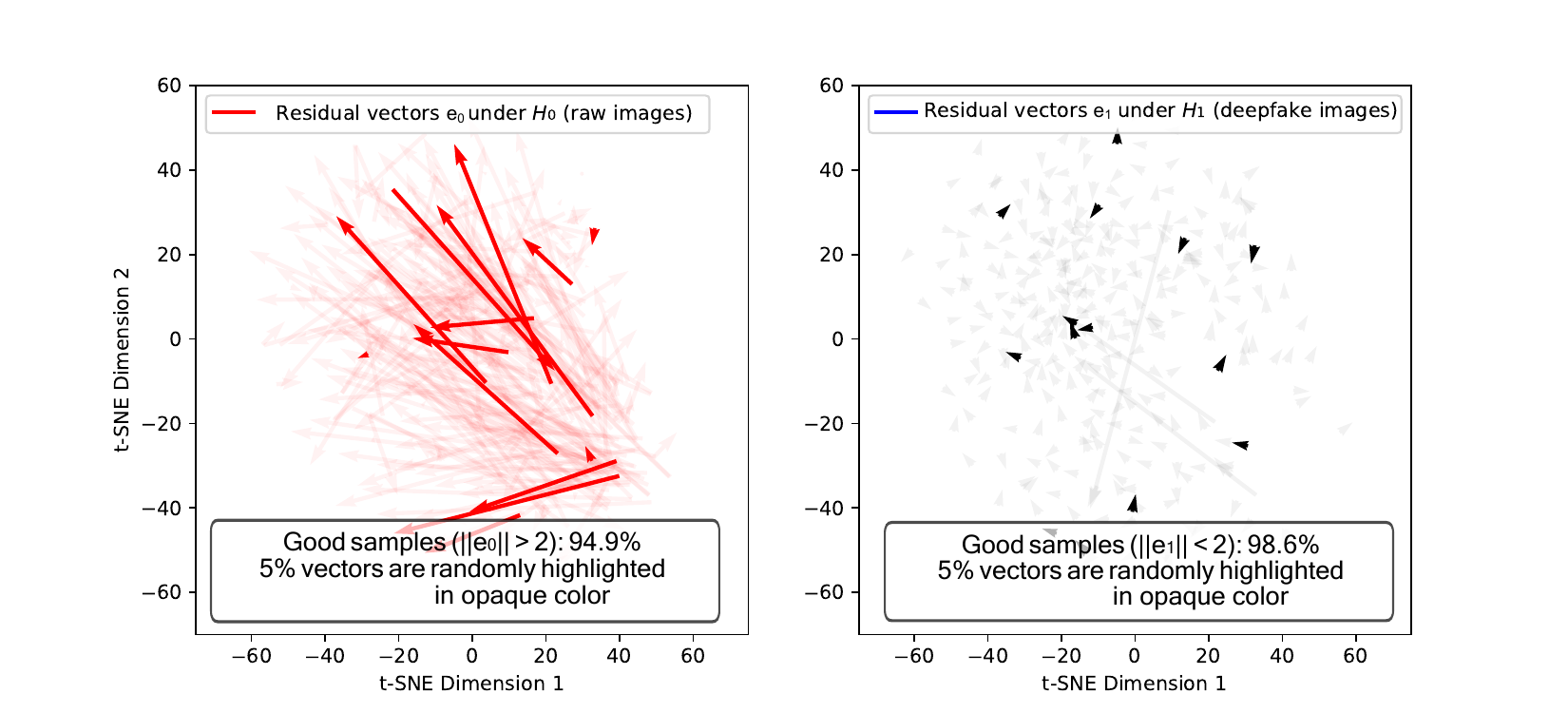}}\vspace{-1mm} 
    \vspace{1ex} 
    \makebox[0.28\textwidth][c]{\small(b)}\vspace{-3.4ex} 
    \makebox[0.75\textwidth][c]{\small(c)} 
        \caption{(a) t-SNE visualization of the feature space of authentic faces from CelebDF \(f\), singly processed images \(\mathrm{FG}(f)\), and doubly processed images \(\mathrm{FG}(\mathrm{FG}(f))\) generated by the learned reconstruction operator using Faceswap-GAN. The sample densities of the first and second classes are more clearly distinguishable than those of the second and third classes. (b)~\(e_0 = \mathrm{FG}(f) - f \) in the t-SNE domain, (c)~\(e_1 = \mathrm{FG}(\mathrm{FG}(f)) - \mathrm{FG}(f) \) in the t-SNE domain. Vectors in (b) are significantly larger compared to those in (c).}
	\vspace{-5mm}
    \label{fig:vector_shifts_celebdf}
\end{figure}

Fig.~\ref{fig:vector_shifts_cacd}\,(a) shows the t-SNE visualization of the feature space for authentic, singly processed, and doubly processed images from the CACD dataset.
(a)~\(e_0 = \mathrm{FG}(f) - f \) and  (b)~\(e_1 = \mathrm{FG}(\mathrm{FG}(f)) - \mathrm{FG}(f) \), when is Faceswap-GAN~(FG)
The visualization reveals significant overlap between the features of singly and doubly processed images, while features from authentic images form a separate distribution. We illustrate the vector residual in the t-SNE feature space concerning a single image, from its authentic state to its singly processed version \(e_0\) in Fig.~\ref{fig:vector_shifts_cacd}\,(b), and from the singly processed version to its doubly processed version \(e_1\) in Fig.~\ref{fig:vector_shifts_cacd}\,(c). These plots highlight the vector residuals in the t-SNE space for the CACD dataset derived from the feature values before applying the Siamese network. This figure shows that the first operation results in significant vector residuals, whereas the second operation leads to minimal residuals for $81.3\%$ of the samples. The t-SNE and residual vector plots for the CelebDF dataset are shown in Fig.~\ref{fig:vector_shifts_celebdf}, with \(94.9\%\) of the residuals being small in the second operation.
This better result arises from the reconstruction operator being trained on a subset of the CelebDF dataset.

\subsubsection{\(R_{\mathrm{df}} = R_{\mathrm{recon}} = \mathrm{D}\)}
To verify the near-idempotence property for diffusion-based deepfake generators, we applied two consecutive diffusion operations to the authentic images. TABLE~\ref{tab:idempotence_summary2} demonstrates that the diffusion-based methods hold the near-idempotence property.
\begin{table*}[!b]
\renewcommand{\arraystretch}{1.2}  
\centering
\caption{Vector Shifting Statistics for two Diffusion Operations.\vspace{-3mm}}
\vspace{2mm}
\scalebox{1.0}
{
\begin{tabular}{|c|cc|cc|}
\hline
\multirow{2}{*}{Dataset} & \multicolumn{2}{c|}{\begin{tabular}[c]{@{}c@{}}Norm of Feature vector residual \(\|e\|\)\\ \textless 7.5, \textless 7.75, \textless 8.0\end{tabular}} & \multicolumn{2}{c|}{\begin{tabular}[c]{@{}c@{}}Norm of Siamese Feature Vector Residual \(\|e\|\)\\ \textless 1.5, \textless 1.6, \textless 1.7\end{tabular}} \\ \cline{2-5} 
                         & \multicolumn{1}{c|}{First Operation \(\|e_0\|\)}                                          & Second Operation \(\|e_1\|\)                                         & \multicolumn{1}{c|}{First Operation \(\|e_0\|\)}                                              & Second Operation \(\|e_1\|\)                                            \\ \hline
CelebDF                  & \multicolumn{1}{c|}{9.74 \%, 13.08 \%, 17.41 \%}                              & 80.22 \%, 85.58 \%, 89.42 \%                             & \multicolumn{1}{c|}{25.35 \%, 29.07 \%, 33.06 \%}                                 & 76.84 \%, 80.80 \%, 84.20 \%                                \\ \hline
CACD                     & \multicolumn{1}{c|}{4.55 \%, 7.00 \%, 10.44 \%}                               & 90.73 \%, 93.18 \%, 95.38 \%                             & \multicolumn{1}{c|}{4.05 \%, 5.36 \%, 6.61 \%}                                    & 70.31 \%, 74.75 \%, 79.44 \%                                \\ \hline
\end{tabular}
}
\label{tab:idempotence_summary2}
\end{table*}
Compared to the results of the Faceswap-GAN shown in TABLE~\ref{tab:idempotence_summary}, diffusion-based operators reveal a larger average norm of the feature vector residual. Thus, slightly higher threshold points were chosen in TABLE~\ref{tab:idempotence_summary3} to better emphasize the differences between the first and second operations in the CDF values. At these selected thresholds, the CDF values for the original feature vector for the first operation range from $4.55\%$ to $17.41\%$, while for the second operation, they range from $80.22\%$ to $95.41\%$.
Notably, the smaller CDF values for the first operation and the larger values for the second operation in the CACD dataset, compared to the CelebDF dataset, suggest that diffusion images generated from CACD images are more easily detected when using the original feature vectors. For the Siamese feature vector, the norm of the residual feature vector ranges from 4.05\% to 33.06\% for the first operation and from 70.31\% to 84.20\% for the second operation. The CDF values approaching $1$ for the second operation again support the near-idempotence property. The PDF plots of feature vector residual norms are shown in the four subfigures of the second column in Fig.~\ref{fig:mainfigure_pdf}.

\begin{figure}[!t]%
    \begin{subfigure}[b]{\linewidth}
        \centering
        \includegraphics[width=1.01\linewidth,trim={0.3cm 0.2cm 1.3cm 1.5cm},clip]{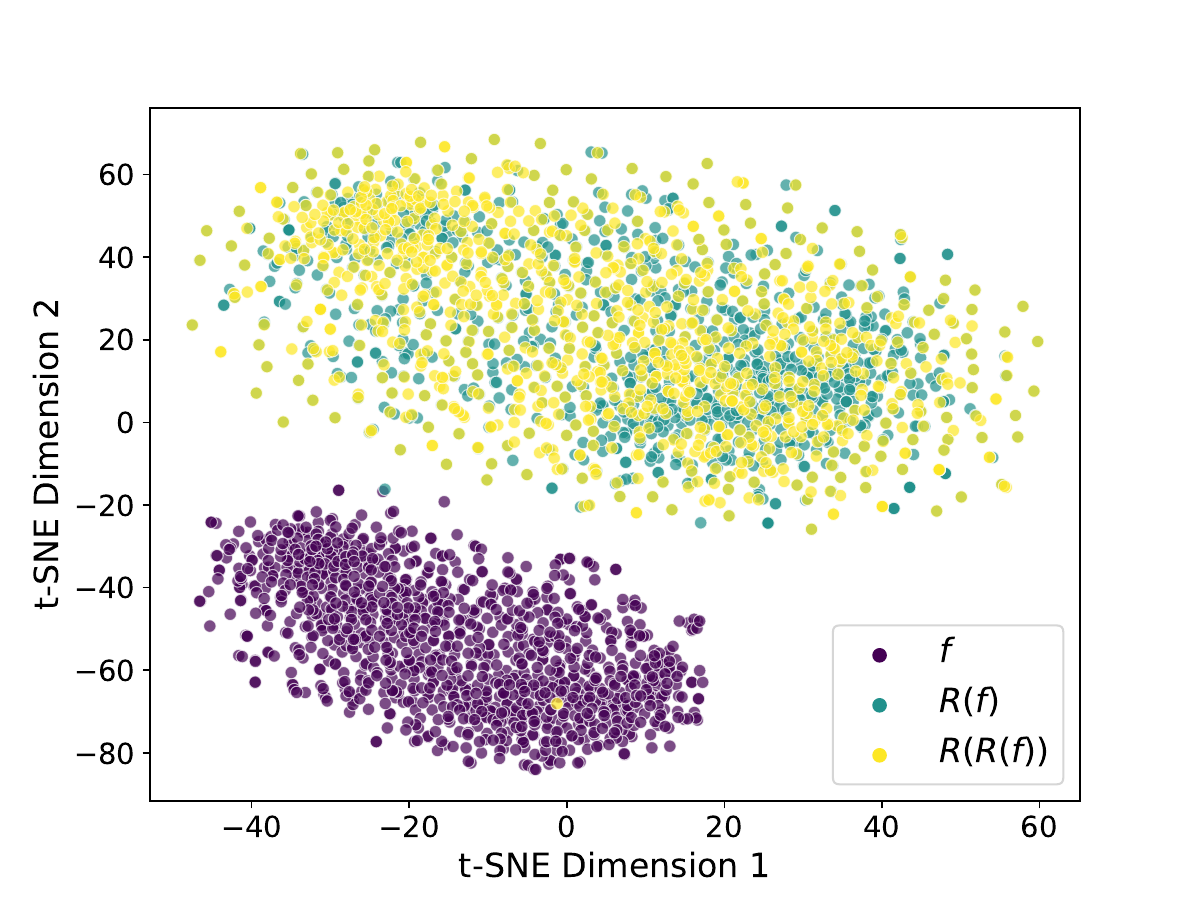}
        \caption{}
        \label{fig_intro2:image1}
        \vspace{0.3cm}
    \end{subfigure}
	\hspace{-2mm}
	{\includegraphics[width=1.01\linewidth,trim={1.5cm 0.1cm 2.4cm 1.5cm},clip]{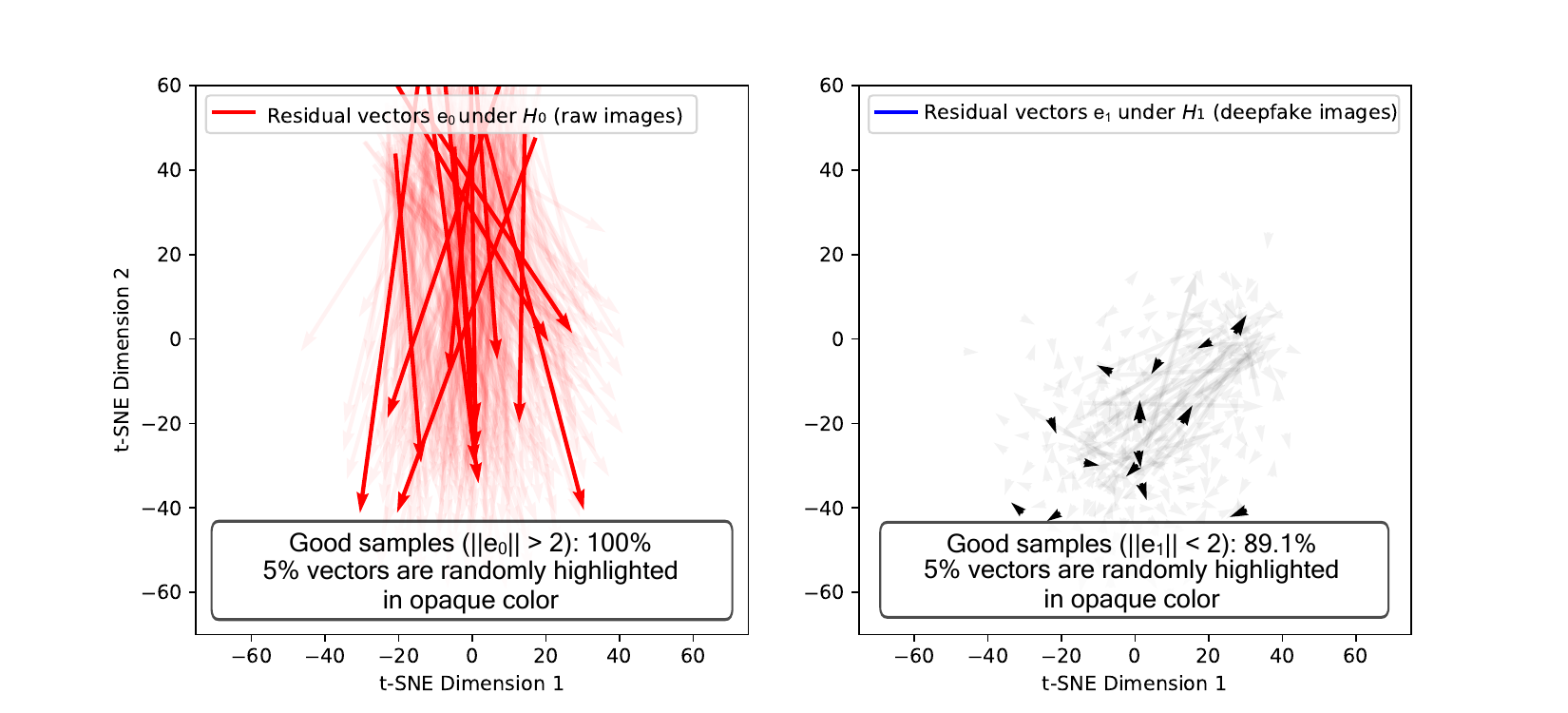}}\vspace{-1mm} 
    \vspace{1ex} 
    \makebox[0.28\textwidth][c]{\small(b)}\vspace{-3.4ex} 
    \makebox[0.75\textwidth][c]{\small(c)} 
        \caption{(a) t-SNE visualization of the feature space of authentic faces from CACD \(f\), faceswapped images generated by DiffSwap first time \(\mathrm{D}(f)\), and second time with repeated operations \(\mathrm{D}(\mathrm{D}(f))\). (b)~\(e_0 = \mathrm{D}(f) - f \) in the t-SNE domain, (c)~\(e_1 = \mathrm{D}(\mathrm{D}(f)) - \mathrm{D}(f) \) in the t-SNE domain.}
	\vspace{-2mm}
    \label{fig:diffswap_shift_diffswap}
\end{figure}

The t-SNE visualizations of the features and vector residual plots are shown in Fig.~\ref{fig:diffswap_shift_diffswap}. In this case, the near-idempotence property is demonstrated by \(89.1\%\) of samples, which is quite an impressive outcome. The significance of this result lies in the fact that the Siamese network was trained using authentic images and deepfake images generated by Faceswap-GAN. Nevertheless, the results in TABLE~\ref{tab:idempotence_summary2} and Fig.~\ref{fig:diffswap_shift_diffswap} show that these features can also work with diffusion-based operations. This indicates that the network is capable of extracting meaningful information from the original feature space, even when applied to a different deepfake generation method.
\subsubsection{\(R_{\mathrm{df}} = \mathrm{D} 
\,\,\&\,\, R_{\mathrm{recon}} = \mathrm{FG}
\)}
In this case, the first operation is diffusion, whereas the second one is Faceswap-GAN. The results for this scenario are presented in TABLE~\ref{tab:idempotence_summary3}.
\begin{table*}[]
\renewcommand{\arraystretch}{1.2}  
\centering
\caption{Vector Shifting Statistics for Diffusion Followed by Faceswap-GAN.\vspace{-3mm}}
\vspace{2mm}
\scalebox{1.0}
{
\begin{tabular}{|c|cc|cc|}
\hline
\multirow{2}{*}{Dataset} & \multicolumn{2}{c|}{\begin{tabular}[c]{@{}c@{}}Norm of Feature vector residual \(\|e\|\)\\ \textless 7.75, \textless 8.0, \textless 8.25\end{tabular}} & \multicolumn{2}{c|}{\begin{tabular}[c]{@{}c@{}}Norm of Siamese Feature Vector Residual \(\|e\|\)\\ \textless 2.1, \textless 2.2, \textless 2.3\end{tabular}} \\ \cline{2-5} 
                         & \multicolumn{1}{c|}{First Operation \(\|e_0\|\)}                                           & Second Operation \(\|e_1\|\)                                         & \multicolumn{1}{c|}{First Operation \(\|e_0\|\)}                                              & Second Operation \(\|e_1\|\)                                            \\ \hline
CelebDF                  & \multicolumn{1}{c|}{12.00 \%, 16.46 \%, 22.73 \%}                              & 75.93 \%, 80.75 \%, 84.93 \%                             & \multicolumn{1}{c|}{54.04 \%, 58.28 \%, 62.61 \%}                                 & 73.85 \%, 77.16 \%, 80.20 \%                                \\ \hline
CACD                     & \multicolumn{1}{c|}{6.35 \%, 9.20 \%, 13.23 \%}                                & 71.01 \%, 75.45 \%, 79.83 \%                             & \multicolumn{1}{c|}{26.35 \%, 30.66 \%, 35.03 \%}                                 & 71.04 \%, 75.49 \%, 79.27 \%                                \\ \hline
\end{tabular}
}
\label{tab:idempotence_summary3}
\end{table*}
The results indicate that even when the second operation differs from the first, the image is only marginally altered for the second operation in the feature space. This supports our proposed system, as a forensic expert may not always know the exact deepfake method used to generate a fake image. The PDF plots of feature vector residual norms are shown in the four subfigures of the last column in Fig.~\ref{fig:mainfigure_pdf}.

\begin{figure}[t!]%
    \begin{subfigure}[b]{\linewidth}
        \centering
        \includegraphics[width=1.01\linewidth,trim={0.3cm 0.2cm 1.3cm 1.5cm},clip]{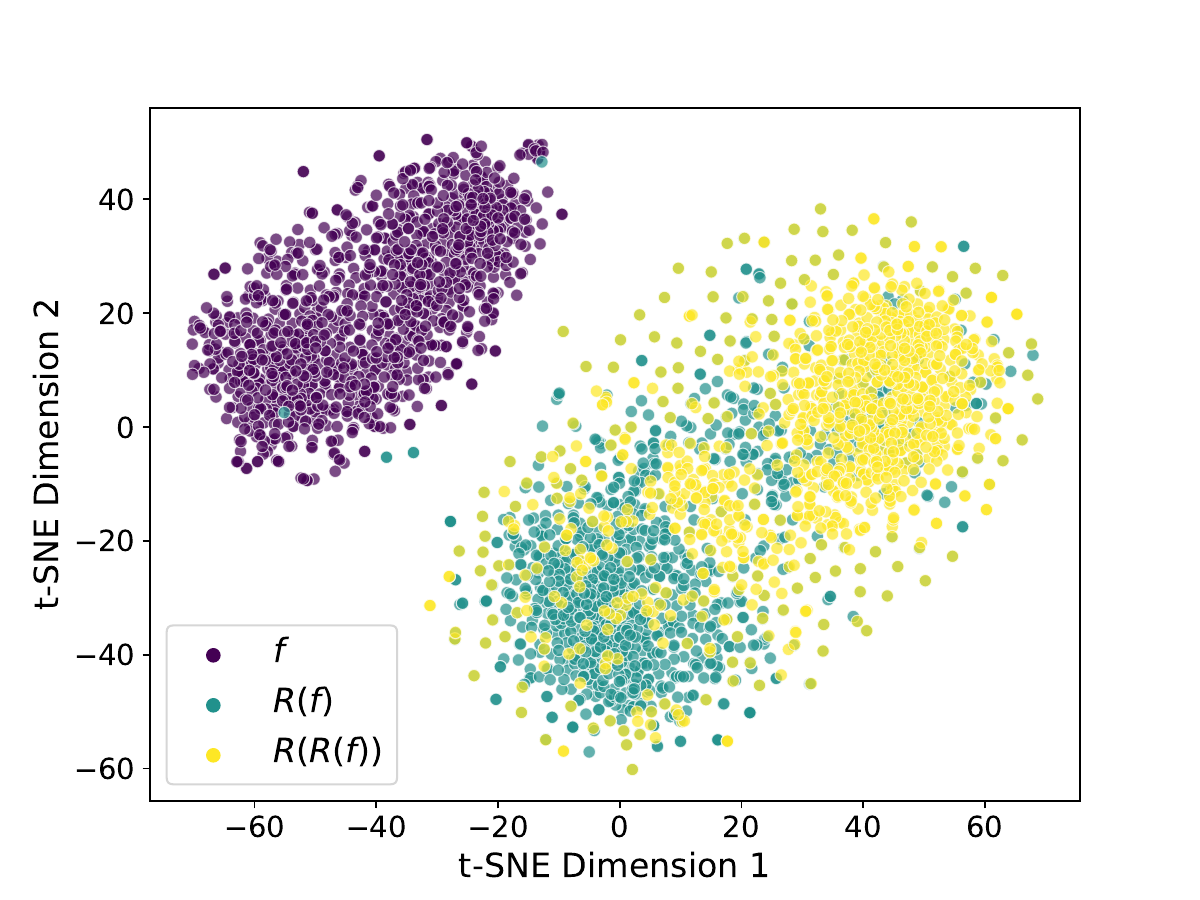}
        \caption{}
        \label{fig_intro2:image1}
        \vspace{0.3cm}
    \end{subfigure}
	\hspace{-2mm}
	{\includegraphics[width=1.01\linewidth,trim={1.5cm 0.1cm 2.4cm 1.5cm},clip]{sec/figs/vec_shift_diff_gan_cacd2.pdf}}\vspace{-1mm} 
    \vspace{1ex} 
    \makebox[0.28\textwidth][c]{\small(b)}\vspace{-3.4ex} 
    \makebox[0.75\textwidth][c]{\small(c)} 
        \caption{(a) t-SNE visualization of the feature space of authentic faces from CACD \(f\), faceswapped images generated by DiffSwap \(\mathrm{D}(f)\), and further processed by FaceSwap-GAN \(\mathrm{FG}(\mathrm{D}(f))\). (b)~\(e_0 = \mathrm{FG}(f) - f \) in the t-SNE domain, (c)~\(e_1 = \mathrm{FG}(\mathrm{D}(f)) - \mathrm{D}(f) \) in the t-SNE domain.}
	\vspace{-2mm}
    \label{fig:diffswap_shift_faceswap}
\end{figure}

However, the feature vector residual in the Siamese feature vector space as shown in Fig.~\ref{fig:diffswap_shift_faceswap} reveals that the Siamese network, trained to distinguish between similar deepfake operators, is less effective when two different types of deepfake operations are involved.
This suggests the need for properly training the Siamese network regarding the types of deepfake operations in both the first and second stages, as testing on different combinations leads to suboptimal results.
\subsection{Summary of the Experimental Results}
Fig.~\ref{fig:vector_shifts_celebdf}\,(a) presents the t-SNE features for the CelebDF dataset, which was used to train reconstruction operators. Notably, the reconstruction operators were trained on a subset of the CelebDF dataset, while the results are presented for the remaining subset. Fig.~\ref{fig:vector_shifts_celebdf}\,(b) and Fig.~\ref{fig:vector_shifts_celebdf}\,(c) illustrate the t-SNE vector residuals resulting from deepfake operations. The large vector residuals in Fig.~\ref{fig:vector_shifts_celebdf}\,(b) indicate a significant change in the image features for the first deepfake operation. On the other hand, minimal vector residuals are observed in Fig.~\ref{fig:vector_shifts_celebdf}\,(c) shows the results when the deepfake operator is applied a second time to the singly processed images. This behavior reflects near-idempotence, supported by $94.9\%$ of the samples. Fig.~\ref{fig:diffswap_shift_diffswap} presents a similar result when both operations are diffusion-based. Once again, in this case, the good samples supporting near-idempotence maintain a high percentage of $89.1\%$. Finally, Fig.~\ref{fig:diffswap_shift_faceswap} shows the third case, when the first operation is diffusion-based and the second operation is Faceswap-GAN. The results indicate that the good samples are reduced to $53.3\%$ in this case. 

\end{document}